# LaAlO$_3$ stoichiometry found key to electron liquid formation at LaAlO$_3$/SrTiO$_3$ interfaces

M. P. Warusawithana[1*], C. Richter[2,3], J. A. Mundy[4], P. Roy[1], J. Ludwig[1], S. Paetel[2], T. Heeg[5], A. A. Pawlicki[1], L. F. Kourkoutis[4,6], M. Zheng[7], M. Lee[1], B. Mulcahy[7], W. Zander[8], Y. Zhu[4], J. Schubert[8], J. N. Eckstein[7], D. A. Muller[4,6], C. Stephen Hellberg[9], J. Mannhart[3], D. G. Schlom[5,6]

[1] National High Magnetic Field Laboratory, Department of Physics, Florida State University, Tallahassee, Florida 32310, USA
[2] Center for Electronic Correlations and Magnetism, University of Augsburg, D-86135 Augsburg, Germany
[3] Max-Planck-Institut für Festkörperforschung, Heisenbergstrasse 1, D-70569 Stuttgart, Germany
[4] School of Applied and Engineering Physics, Cornell University, Ithaca, New York 14853, USA
[5] Department of Materials Science and Engineering, Cornell University, Ithaca, New York 14853, USA
[6] Kavli Institute at Cornell for Nanoscale Science, Ithaca, New York 14853, USA
[7] Department of Physics, University of Illinois at Urbana-Champaign, Urbana, Illinois 61801, USA
[8] Peter Gruenberg Institute 9, JARA-Fundamentals of Future Information Technologies, Research Centre Jülich, D-52425 Jülich, Germany
[9] Center for Computational Materials Science, Naval Research Laboratory, Washington, DC 20375, USA
*e-mail: maitri@magnet.fsu.edu

## Abstract

Emergent phenomena, including superconductivity and magnetism, found in the two-dimensional electron liquid (2-DEL) at the interface between the *insulators* LaAlO$_3$ and SrTiO$_3$ distinguish this rich system from conventional two-dimensional electron gases at compound semiconductor interfaces. The origin of this 2-DEL, however, is highly debated with focus on the role of defects in the SrTiO$_3$ while the LaAlO$_3$ has been assumed perfect. Our experiments and first principles calculations show that the cation stoichiometry of the nominal LaAlO$_3$ layer is key to 2-DEL formation: only Al-rich LaAlO$_3$ results in a 2-DEL. While extrinsic defects including oxygen deficiency are known to render LaAlO$_3$/SrTiO$_3$ samples conducting, our results show that in the absence of such extrinsic defects, an interface 2-DEL can form. Its origin is consistent with an intrinsic electronic reconstruction occurring to counteract a polarization catastrophe. This work provides a roadmap for identifying other interfaces where emergent behaviors await discovery.

The intriguing discovery of a two-dimensional electron liquid (2-DEL)[1-3] at the interface between two *simple band insulators*, lanthanum aluminate (LaAlO$_3$) and strontium titanate (SrTiO$_3$), and the subsequent observation of collective ground states including superconductivity[4,5], magnetism[6,7] and an unexpected coexistence of superconductivity and magnetism[8-10], has fueled exciting research on the interactions between the confined electrons and their spin, orbital degrees of freedom[11-13]. While the origin of this 2-DEL has been attributed to a "*polar catastrophe*," an intrinsic electronic reconstruction[1,14-18], which acts to remove the diverging electric potential caused by the atomic layer arrangement at the interface – (SrO)$^0$-(TiO$_2$)$^0$-(LaO)$^{+1}$-(AlO$_2$)$^{-1}$ – there has been much debate as to the role of defects on the conductivity of this system[19-23]. Even though a growing number of experiments have provided evidence supporting the *polar catastrophe* mechanism[24,25], the relatively high-energy and far-from-equilibrium growth conditions involved in pulsed-laser deposition (PLD), which has been at the forefront of the aforementioned discoveries, has raised concerns about defect-driven conductivity at the interface[22,23]. As the effect of the LaAlO$_3$ composition on interfacial conductivity has not been studied[26], we study that here on samples mostly grown by molecular-beam epitaxy (MBE). Our experiments not only eliminate suggested extrinsic defect mechanisms as the origin of the interface conductivity, but show an unexpected dependence on the cation stoichiometry of the LaAlO$_3$ film for the formation of an interfacial 2-DEL. Our results reveal that a La/Al ratio ≤ 0.97±0.03 is a necessary condition for obtaining a 2-DEL at the interface between LaAlO$_3$ and TiO$_2$-terminated (100) SrTiO$_3$. Our studies further show that this result is also consistent with samples grown by PLD[2].

**Results**

Mosaic growth experiment

To investigate the effect of LaAlO$_3$ stoichiometry on the 2-DEL as well as the role of substrate preparation, growth conditions, and intermixing at the interface, we grew simultaneously on a mosaic of substrates mounted on a sample holder (Fig. 1a), with a monotonic variation in the La/Al ratio across the LaAlO$_3$ film achieved from the placement of the sources in the MBE system. See methods. These 16 substrates were obtained from cutting four, 10 mm x 10 mm, TiO$_2$-terminated (100) SrTiO$_3$ substrates (labeled 1, 2, 3 and 4) each into four 5 mm x 5 mm pieces (labeled A, B, C and D). The mosaic was formed by mounting pieces from different substrates next to each other. With lanthanum and aluminum fluxes calibrated to within a few percent, the gradient in the La/Al ratio ensured that at some position on the mosaic, stoichiometric LaAlO$_3$ was deposited. It also provided a means to explore the effect of small composition variations in La$_{(1-\delta)}$Al$_{(1+\delta)}$O$_3$ on the resulting electronic properties at the interface. This growth experiment was repeated three times. In each growth, the position on the mosaic where stoichiometric LaAlO$_3$ was expected was moved by adjusting the shutter open times of lanthanum and aluminum for the growth of each monolayer based on the *in situ* flux calibration that preceded the mosaic growth. See Supplementary Discussion 1. Immediately following each growth, a companion calibration sample was also grown on which *ex situ* Rutherford backscattering spectroscopy (RBS) measurements were performed to determine the position of stoichiometric $LaAlO_3$. See Supplementary Discussion 2. This position is indicated for each mosaic in Fig. 1a by a white circle. Moving to the left of the white circle increases the La/Al ratio, given by $(1-\delta)/(1+\delta)$, beyond 1.00, while moving to the right decreases the La/Al ratio below 1.00.

The interfacial electronic properties of each mosaic sample were probed with contacts made to the interfaces patterned using photolithography. See methods for patterning details. In all three mosaic experiments, conducting interfaces (indicated in green in Fig. 1a) were found only in samples that were mounted to the right of the white circle, where La/Al < 1.00, regardless of the substrate from which the samples originated. Figure 1b shows the temperature dependence of the resistance of many of the conducting interfaces obtained from the mosaic experiments. A residual resistivity ratio (resistance at 290K/resistance at 4K) of the order of 50 is observed in these conducting samples, which are also found to be superconducting with superconducting critical temperatures between 150 mK and 250 mK. A representative superconducting transition is shown in Fig. 1c. The conductivity and superconductivity observed in these samples are comparable to those reported in PLD-grown samples[2,4,5].

Evidence eliminating extrinsic mechanisms of conduction
This mosaic growth experiment eliminates four of the proposed extrinsic causes of conductivity in $LaAlO_3$/(100) $SrTiO_3$ samples: (1) oxygen vacancies in the $SrTiO_3$ due to insufficiently oxidizing conditions during growth or during an oxygen annealing step after growth[19-21,27], (2) high concentrations of oxygen vacancies in the surface region of etched and annealed $SrTiO_3$ substrates prepared by the preferred substrate termination etch for (100) SrTiO$_3$[28], (3) bombardment of the SrTiO$_3$ by energetic species during growth leading to oxygen vacancies[29], and (4) chemical mixing of lanthanum from the LaAlO$_3$ with the SrTiO$_3$ to make La-doped SrTiO$_3$[30], a known conductor and superconductor. All of the SrTiO$_3$ pieces (cut from 4 different substrates) were grown on at the same time by MBE. If the growth conditions were insufficient to oxidize one of them, the other pieces cut from the same substrate need also be conducting, thus eliminating hypothesis (1). Similarly, hypothesis (2) can be eliminated as all SrTiO$_3$ pieces were prepared using the same TiO$_2$-termination procedure for (100) SrTiO$_3$[31], but a 2-DEL was found only in some pieces of each substrate.

In contrast to PLD growth, where conditions may exist for high-energy species to bombard the growing film and produce various types of extrinsic defects[32-34], MBE affords a gentle means of film growth utilizing only thermal beams of neutral species, including the purified ozone beam used as the oxidant, with energies <<1 eV. This eliminates hypothesis (3). Finally, LaAlO$_3$ is in contact with SrTiO$_3$ on all of the simultaneously grown wafers, so if intermixing (hypothesis (4)) is relevant, the intermixing is only active in samples containing certain compositions. All three mosaic experiments that cover an even larger composition space show that La-rich compositions – where the most interdiffusion might be expected – are, in fact, not conducting.

The interfaces of both conducting and non-conducting samples from the same mosaic growth were chemically mapped using cross-sectional scanning transmission electron microscopy with electron energy loss spectroscopy (STEM-EELS). The high angle annular dark field (HAADF) STEM images in Fig. 2 show coherent LaAlO$_3$/SrTiO$_3$ interfaces of representative insulating and conducting interfaces near the ends of the stoichiometry range explored. EELS spectroscopic imaging revealed a small amount of lanthanum interdiffusion (<0.1%) into the substrate for both conducting and insulating samples (Fig. 2c). Critically the lanthanum interdiffusion was not correlated to the $La_{(1-\delta)}Al_{(1+\delta)}O_3$ stoichiometry or the conductivity, ruling out hypothesis (4). Fine structure analysis of the O-$K$ edge showed a distinct interface component with the same concentration and spatial extent for both conducting and insulating samples (Supplementary Fig. S3). Changes in the titanium valence were below the sensitivity threshold of the microscope (Supplementary Fig. S4). Further, from measurements made on plan view samples, we find the density and spacing of interfacial dislocations at the $SrTiO_3/LaAlO_3$ interface to be comparable (~38±3 nm vs. 43±3 nm between dislocations for conducting and insulating samples 2-4B and 2-4A, respectively).

Effect of $La_{(1-\delta)}Al_{(1+\delta)}O_3$ stoichiometry on 2-DEL formation

To more accurately and quantitatively determine the effect of $LaAlO_3$ stoichiometry on 2-DEL formation, we grew on (100) $SrTiO_3$ crystals that were ~25 mm long (see methods). Immediately following the growth of a calibration sample (see Supplementary Discussion 1), three such long samples were grown back-to-back. The La/Al ratio at the center of each sample was varied by adjusting the shutter-open-times of lanthanum and aluminum sources for the growth of each monolayer. Immediately following the growth of the three samples, a companion RBS calibration sample was also grown. Figure 3a shows a representative RBS spectrum from a companion calibration sample. The La/Al ratio at the center of each long sample was determined from peak integration of the RBS spectrum. The relative spatial variation of the La/Al ratio was also determined to better than ±3% accuracy through separate RBS experiments. See Supplementary Discussion 2. From these RBS measurements, the absolute spatial variation of the La/Al ratio of the three long $La_{(1-\delta)}Al_{(1+\delta)}O_3$ /SrTiO$_3$ samples was determined.

The long $La_{(1-\delta)}Al_{(1+\delta)}O_3$ /SrTiO$_3$ samples were patterned (see methods) to enable 4-point transport measurements every 1.0 mm along the ~25 mm length of the samples (Fig. 3b). This 1 mm spatial separation roughly corresponded to a 1.0% change in the La/Al ratio. Room temperature resistance measurements made on these three samples are plotted as a function of the La/Al ratio (Fig. 3c and Supplementary Fig. S5). Note that the three long samples are offset from each other in their span of La/Al ratio. The transport data of all three samples are, however, consistent with a sharp insulator-to-metal transition that occurs at La/Al ratio, $(1-\delta)/(1+\delta) \leq 0.97\pm0.03$. This transition is also consistent with the multiple experiments involving a mosaic spread of samples (Fig. 1a).

The composition of a LaAlO$_3$ sample grown in the same PLD system and under similar growth conditions as the samples with conducting interfaces studied in reference 2 was also measured. This RBS measurement yielded a spectrum similar to that shown in Fig. 3a. An arrow in Fig. 3c indicates the stoichiometry of the PLD-grown sample. This measurement further supports the result that La/Al $\leq 0.97\pm0.03$ is a necessary condition for obtaining a 2-DEL at $La_{(1-\delta)}Al_{(1+\delta)}O_3$/SrTiO$_3$ interfaces and suggests that this result is independent of growth method used as long as the growth conditions do not promote extrinsic defects.

Density functional calculations

The $La_{(1-\delta)}Al_{(1+\delta)}O_3$ film accommodates different La/Al ratios by forming defects. We performed density functional calculations to determine the energetically favored defects as a function of La/Al ratio (see Supplementary Discussion 3). Supercells of bulk $LaAlO_3$ strained to the $SrTiO_3$ lattice constant were used. For La/Al ratios below 1, the excess aluminum substitutes for lanthanum. Each substitutional $Al^{3+}$ displaces from the ideal $La^{3+}$ position to bond with five oxygen ions (Fig. 4a and Supplementary Fig. S13), resulting in a dipole moment but no net charge. For La/Al ratios above 1, the excess lanthanum cannot substitute for aluminum, and $Al_2O_3$-vacancy-complexes form as shown in Fig. 4d. Larger $Al_2O_3$-vacancy-complexes (see Supplementary Fig. S6) relieve more strain in the films and have lower energy.

The large electric field density in ideal films of $LaAlO_3$ on $TiO_2$-terminated (001) $SrTiO_3$ has an energy cost. Negative charges at the interface screen this field, and indeed density functional calculations simulating thick films of $LaAlO_3$ on $SrTiO_3$ show the interface is unstable to the formation of cation vacancies. For La/Al > 1, the $Al_2O_3$-vacancy-complexes provide conduits for cations to move from the interface. With the positive charge associated with the cations that have moved away from the interface missing, the field is screened and the interface remains insulating (see, for example, Fig. 4e,f). In contrast, for La/Al ≤ 1, the aluminum substitutional defects block migration of cations from the interface to form vacancies, resulting in electronic charge to transfer from the surface to the interface and form a 2-DEL to avoid a polar catastrophe (Fig. 4b,c). It is likely that a surface reconstruction accompanies the electronic charge transfer to the interface – density functional calculations have shown oxygen vacancies on the surface have negative formation energy at LaAlO$_3$ thicknesses greater than 3 unit cells[35]. There will be local fluctuations, however, in the stoichiometry of the grown films. For example, films grown on stoichiometry will have Al-rich regions and La-rich regions. We suspect that a percolating network of these local stoichiometry variations shifts the critical La/Al ratio to the smaller value (0.97±0.03) seen in the experiments.

Cation composition mapping using STEM-EELS

To assess the above theory predicting different types of defects in films with La/Al > 1 from those with La/Al < 1, we utilized atomic-resolution STEM-EELS to determine layer-to-layer variations in cation occupancy along the [001] growth direction. Spectroscopic images of lanthanum and strontium as well as aluminum and titanium were collected simultaneously to map the total concentration of cations residing on the *A*-site and *B*-site sublattices, respectively. See Supplementary Discussion 4 for details. In the conducting, Al-rich samples, aluminum and titanium were found to completely occupy the *B*-site sublattice, as shown in Fig. 5b. In contrast, the insulating, La-rich samples show a local accumulation of cation vacancies on the *B*-site sublattice at the interface (Fig. 5d). Figure 5b,d also shows that neither interface shows a corresponding local dip in the cation concentration on the *A*-site sublattice. The experimentally observed cation vacancies in the *B*-site sublattice at the interface of La-rich samples (Fig. 5d) are significant for understanding the charge balance at the interface. Density functional calculations show that these cation vacancies alleviate the polar catastrophe (Fig. 4f and Fig. 5c) without the transfer of electronic charge to the interface to form a 2-DEL.

**Discussion**

It is well known that extrinsic defects can lead to conductivity in SrTiO$_3$ and in LaAlO$_3$/(100) SrTiO$_3$ samples. The question has been whether there could be an intrinsic mechanism, such as the *polar catastrophe*, that could lead to conductivity at the interface between the two band insulators. Our uniquely implemented growth experiments involving over 50 samples together with careful stoichiometry measurements have answered this question by comparing conductive and insulating interfaces grown simultaneously, even on the same substrate. Minor differences in the stoichiometry of the La$_{(1-\delta)}$Al$_{(1+\delta)}$O$_3$ films correlate to the abrupt change from conducting to insulating samples at a critical La/Al ratio of 0.97±0.03. This result, consistent with the critical thickness found[2,25] in samples grown over a large span of oxygen concentrations, further supports that oxygen vacancies cannot be the driver of conductivity (also see Supplementary Discussion 5). The ability to compare insulating and conducting interfaces in our sample set has enabled us to assess which characteristics correlate with conductive interfaces and which do not. None of the proposed extrinsic mechanisms – (1) oxygen vacancies in the SrTiO$_3$ due to insufficiently oxidizing conditions[19-21,27], (2) high concentrations of oxygen vacancies in the surface region of etched and annealed SrTiO$_3$ substrates[28], (3) bombardment of the SrTiO$_3$ by energetic species during growth[29], or (4) chemical mixing of lanthanum from the LaAlO$_3$ with the SrTiO$_3$[30] – can explain the sharp transition from conducting to insulating interfaces in our sample set. This is not to say that the above mechanisms cannot lead to conductive samples, they easily can; it is only to say that in our sample set, some other mechanism is at work to make the Al-rich interfaces conductive.

Having ruled out all proposed extrinsic defect mechanisms for the observed 2-DEL in our samples, we have considered how it is that a small change in the $La_{(1-\delta)}Al_{(1+\delta)}O_3$ film composition could lead to the abrupt change in interface conductivity. Our direct spectroscopic observation of the accumulation of cation vacancies near the interface in La-rich samples and its correlation with insulating interfaces, introduces a new factor relevant to 2-DEL formation – cation vacancies. By measuring the *B*-site vacancy content across both conductive and insulating $LaAlO_3/SrTiO_3$ interfaces (grown simultaneously, side-by-side on pieces of the same substrate), we find the insulating interfaces consistently show *B*-site vacancies while the conducting interfaces do not. Our observations, supported by first-principles calculations, point to different pathways for compensating the polar discontinuity at the heterointerface: in La-rich samples cation vacancies form at the interface to avoid the polarization catastrophe which lead to insulating interfaces, while in Al-rich samples the diffusion of cations away from the interface is blocked and an electronic reconstruction occurs to counteract the polarization catastrophe giving rise to a 2-DEL.

Our experiments thus show that a 2-DEL can form at the interface between $La_{(1-\delta)}Al_{(1+\delta)}O_3$ and $TiO_2$-terminated (100) $SrTiO_3$ substrates in the absence of any of the previously proposed extrinsic defect mechanisms that are known to cause conductivity in bulk $SrTiO_3$. Further, our results are consistent with the polar catastrophe model; in insulating (La-rich) samples, the predicted and observed local accumulation of cation vacancies on the *B*-site in the vicinity of the interface acts to remove the diverging potential while this is accomplished in conducting (Al-rich) samples by an electronic reconstruction that forms a 2-DEL. The different defect responses deduced by this work to be operative in La-rich and Al-rich films grown on $TiO_2$-terminated (100) $SrTiO_3$ shows that controlling the composition of the overlying $La_{(1-\delta)}Al_{(1+\delta)}O_3$ is key to 2-DEL formation. This result likely extends to other interfaces with mismatched polarity and lattice strain and is thus important for obtaining 2-DELs as well as other functional properties at buried oxide interfaces including systems that do not involve $SrTiO_3$.

**Methods**

To explore whether the conductivity at the LaAlO$_3$-SrTiO$_3$ interface is generated by PLD-specific defects as suggested[22,23,29], we grew LaAlO$_3$-SrTiO$_3$ interfaces by MBE. In contrast to PLD where a multicomponent target and molecular oxygen are typically used, the oxide heterostructures grown via MBE utilized separate elemental sources of the constituents including distilled ozone as the oxidant. The geometry between the effusion cells and the substrate in our MBE experiments was selected to produce flux gradients of lanthanum and aluminum in approximately opposite directions across the sample. Because the sample was not rotated during growth, the La/Al ratio across the La$_{(1-\delta)}$Al$_{(1+\delta)}$O$_3$ film could be monotonically varied while ensuring that at some position on the sample stoichiometric LaAlO$_3$ was deposited, provided the lanthanum and aluminum fluxes were calibrated to within a few percent (see Supplementary Discussion 1 for *in situ* flux calibration details).

All SrTiO$_3$ substrates were prepared to have a TiO$_2$-terminated surface[31]. The LaAlO$_3$ films were grown in a background partial pressure of distilled ozone of 1×10$^{-6}$ torr at a substrate temperature of 680 ºC in a Veeco 930 MBE. The LaAlO$_3$ films were grown starting with a LaO monolayer followed by an AlO$_2$ monolayer. This monolayer sequence was repeated to obtain an eight-unit-cell-thick film of LaAlO$_3$. The growth rate was kept at approximately 60 s per monolayer. After growth the samples were cooled to below 200 ºC while maintaining the same background partial pressure of distilled ozone. SrTiO$_3$ substrates, when subjected to these same growth conditions except for the deposition of the La$_{(1-\delta)}$Al$_{(1+\delta)}$O$_3$, did not show any signs of conductivity. See Supplementary Discussion 5. Electrical contacts were patterned utilizing photolithography to contact the interface 2-DEL by ion milling the contact areas with argon followed by titanium and gold deposition[2]. Contacts to the mosaic samples were made in a square geometry around the edge of the samples. For the long samples a mask was used to enable local 4-point measurements spaced 1.0 mm apart as shown in Fig. 3b. HAADF-STEM images and EELS spectroscopic images were recorded from cross-sectional TEM specimens using a 100 keV NION UltraSTEM.

**Acknowledgements**
The authors acknowledge discussions and interactions with Scott Chambers, Noam Bernstein, J. Grazul, J. Aarts, J. Brooks and E. J. Kirkland. A portion of this work was performed at the National High Magnetic Field Laboratory, which is supported by National Science Foundation Cooperative Agreement No. DMR-0654118, the State of Florida, and the U.S. Department of Energy. In addition, the financial support of the Deutsche Forschungsgemeinschaft (TRR 80), European Community (OxIDes), the National Science Foundation through the MRSEC program (DMR-1120296), NSF graduate fellowship DGE-0707428 (J.A.M.), the D.O.D. through an NDSEG fellowship (J.A.M.) and Army Research Office ARO #W911NF0910415 is gratefully acknowledged. Computations were performed at the ERDC DoD Major Shared Resource Center.

**Figure 1. Flux gradients and interfacial conductivity of mosaic samples, including superconductivity.** (**a**) Mosaic arrangement of substrates for each growth. The location of lanthanum and aluminum effusion cells relative to the mosaic of substrates is also indicated. La/Al ratio decreases on moving from left to right across each mosaic. The position of stoichiometric $LaAlO_3$, determined by *ex situ* RBS measurements for each mosaic growth, is shown by a white circle. Samples with a conducting interface are shown in green and are only found to the right of the white circle in each mosaic. (**b**) Temperature dependence of resistance of the 2-DEL is plotted for a representative set of conducting samples from the mosaic growths. The samples are labeled with the mosaic number followed by substrate number - for example, 2 – 1D indicates mosaic 2, piece D of substrate 1. (**c**) A representative low-temperature resistance vs. temperature plot shows the 2-DEL is superconducting.

**Figure 2. HAADF-STEM images and EELS spectroscopic maps on conducting and insulating mosaic samples.** (**a**) An insulating sample 2 – 4A (with La/Al=1.06±0.03). (**b**) A conducting sample 2 – 4B (with La/Al=0.90±0.03). The HAADF-STEM images show that both samples have coherent interfaces. The EELS spectroscopic images map the concentration of lanthanum in magenta and titanium in turquoise; both samples show a small amount of interdiffusion at the interface. (**c**) EELS maps of lanthanum for several samples from Mosaic 2 are ordered from left to right by the increasing degree of interdiffusion. No correlation between lanthanum interdiffusion and interface conductivity is observed. Slight distortions in the maps are due to drift and charging during acquisition.

**Figure 3. Interfacial conductivity and its dependence on La/Al ratio of La$_{(1-\delta)}$Al$_{(1+\delta)}$O$_3$ films.** (**a**) Representative RBS spectrum from a companion sample. Well-separated peaks obtained, especially for aluminum, due to the structure of the companion RBS samples, enabled accurate calibration measurements by peak integration. See Supplementary Discussion 2. (**b**) Patterning of a ~25 mm long La$_{(1-\delta)}$Al$_{(1+\delta)}$O$_3$ / (100) SrTiO$_3$ sample allowed local 4-point transport measurements to be made at 1.0 mm intervals. Note that the shadows of the Ti/Au contacts are visible through the transparent sample. (**c**) Room temperature sheet resistance of La$_{(1-\delta)}$Al$_{(1+\delta)}$O$_3$ / (100) SrTiO$_3$ interfaces obtained by local 4-point resistance measurements is plotted as a function of the La/Al ratio determined by RBS measurements. A sharp jump in sheet resistance is observed at La/Al = 0.97±0.03, consistent in all three samples. See Supplementary Fig. S5 for resistance of conducting devices and Supplementary Discussion 2 for error analysis. An arrow indicates the stoichiometry of a PLD grown companion sample similar to the samples studied in reference 2.

**Figure 4. Lowest energy structures determined with density functional theory and illustrations of the polar catastrophe for Al-rich (a-c) and La-rich (d-f) La$_{(1-\delta)}$Al$_{(1+\delta)}$O$_3$ films on SrTiO$_3$.** (**a**) In Al-rich films, aluminum substitutes for lanthanum and shifts off center. The lowest-energy structure is shown as viewed along the [100] direction. (**b**) The alternating charges ($\rho$) of the (001) planes in Al-rich La$_{(1-\delta)}$Al$_{(1+\delta)}$O$_3$ and the charge neutral (001) planes in SrTiO$_3$ generates a positive average electric field ($E$) and a diverging potential ($V$). Note: $\rho/\varepsilon = \nabla \cdot E = -\nabla^2 V$. The substitution of Al$^{3+}$ for La$^{3+}$ does not modify the alternating polarity from that of a

stoichiometric LaAlO$_3$ film. (**c**) In thick Al-rich films, the system reconstructs electronically, transferring ½ electron per unit cell from the surface to the interface. (**d**) In La-rich films, Al$_2$O$_3$-vacancy-complexes form, which are periodic in the [001] direction. The smallest Al$_2$O$_3$-vacancy-complex is shown as viewed along the [001] direction. (**e**) The extended Al$_2$O$_3$-vacancy-complexes in the unreconstructed La-rich films also remove oxygen from the nominal (LaO)$^+$ layers. The aluminum deficiency is given by $x = -2\delta/(1-\delta)$. A diverging potential results in the unreconstructed films from the alternately charged (001) planes of La$_{(1-\delta)}$Al$_{(1+\delta)}$O$_3$. (**f**) In thick La-rich films, extra aluminum vacancies can move to the interface through the Al$_2$O$_3$-vacancy-complexes to screen the diverging potential. The aluminum deficiency $y$ now depends on the stoichiometry ($\delta$) and the film thickness.

**Figure 5. Cation concentrations determined by theory and atomic-resolution STEM-EELS analysis of Al-rich (a-b) and La-rich (c-d) La$_{(1-\delta)}$Al$_{(1+\delta)}$O$_3$ films on SrTiO$_3$.** (**a**) *A*-site and *B*-site cation occupancies found for Al-rich structure shown in Fig. 4c with $\delta=0.05$. Although aluminum substitutional defects are included as part of the *A*-site occupancy, they are shifted off center as shown in Fig. 4a. (**b**) A representative STEM-EELS spectroscopic image of aluminum (pink) and titanium (turquoise) from the conducting wafer 2 – 4B shown in Fig. 2b. The normalized aggregate concentration of titanium and aluminum occupying the *B*-site sublattice does not show any systematic variation from layer to layer along the [001] direction. Similarly, the aggregate occupancy of lanthanum and strontium at the *A*-site shows no variation along the [001] direction. (**c**) *A*-site and *B*-site cation occupancies found for La-rich structure shown in

Fig. 4f with y=0.04. Note the cation deficiency in the *B*-site at the interface. (**d**) A representative STEM-EELS spectroscopic image of aluminum (pink) and titanium (turquoise) from the insulating wafer 2 – 4A shown in Fig. 2a. In contrast to the map shown in (b), there is a dip in the normalized *B*-site occupancy at the interface, indicating a local accumulation of *B*-site vacancies while the *A*-site occupancy shows no detectable variation along the [001] direction.

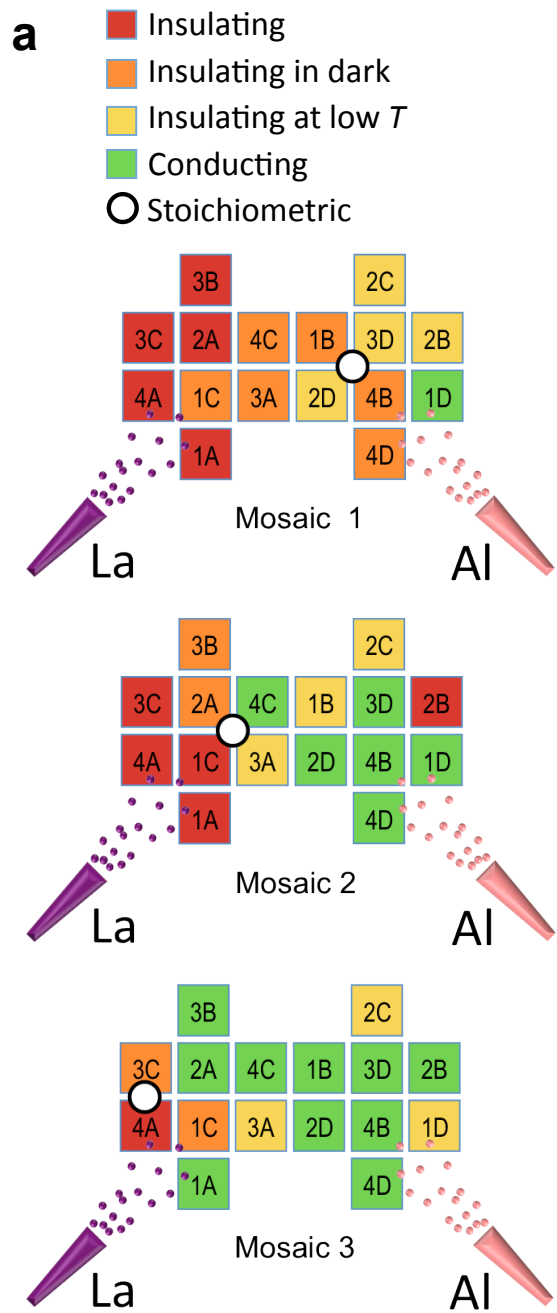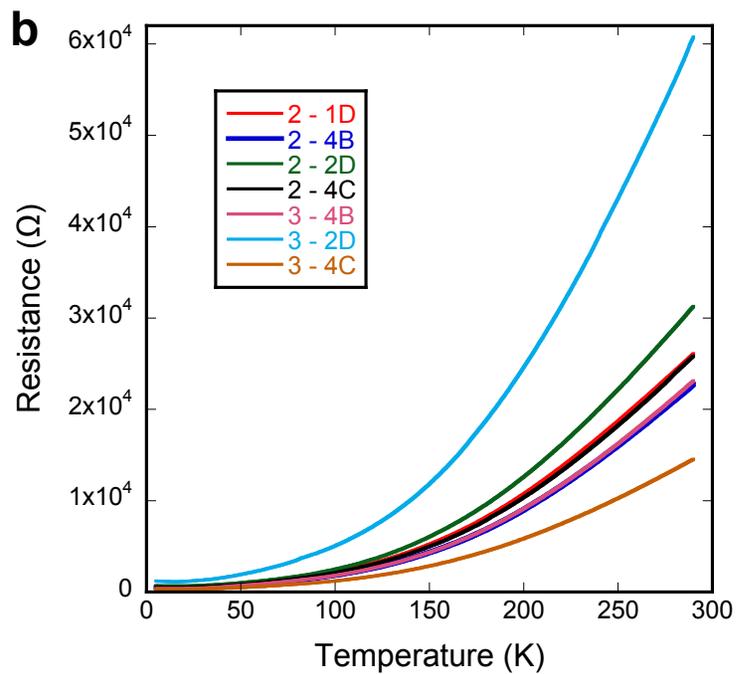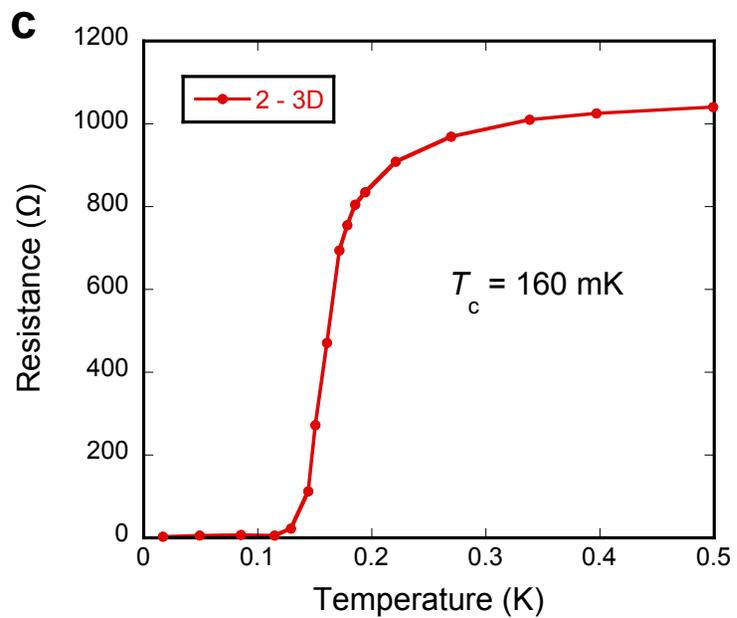

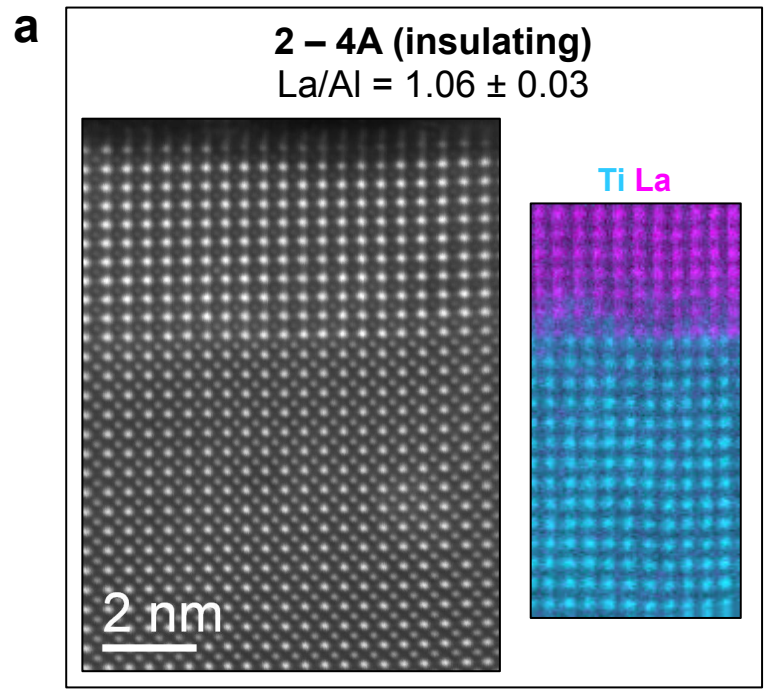
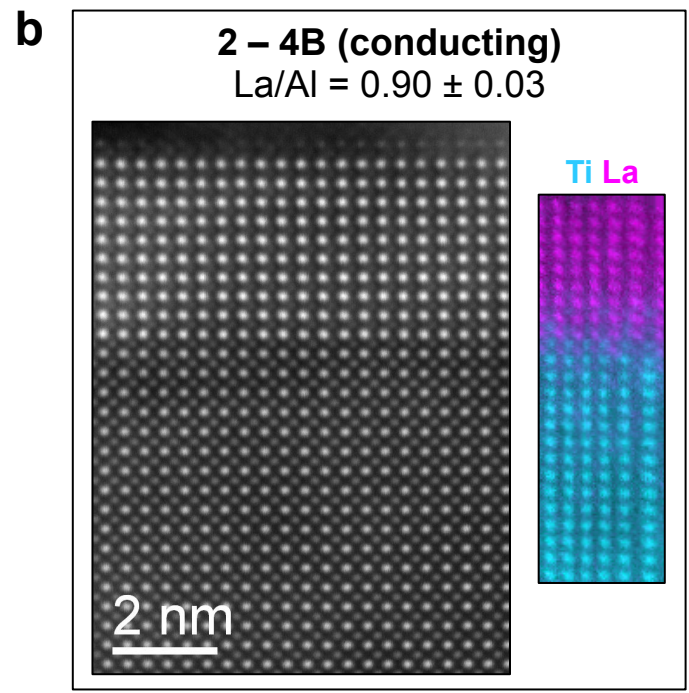
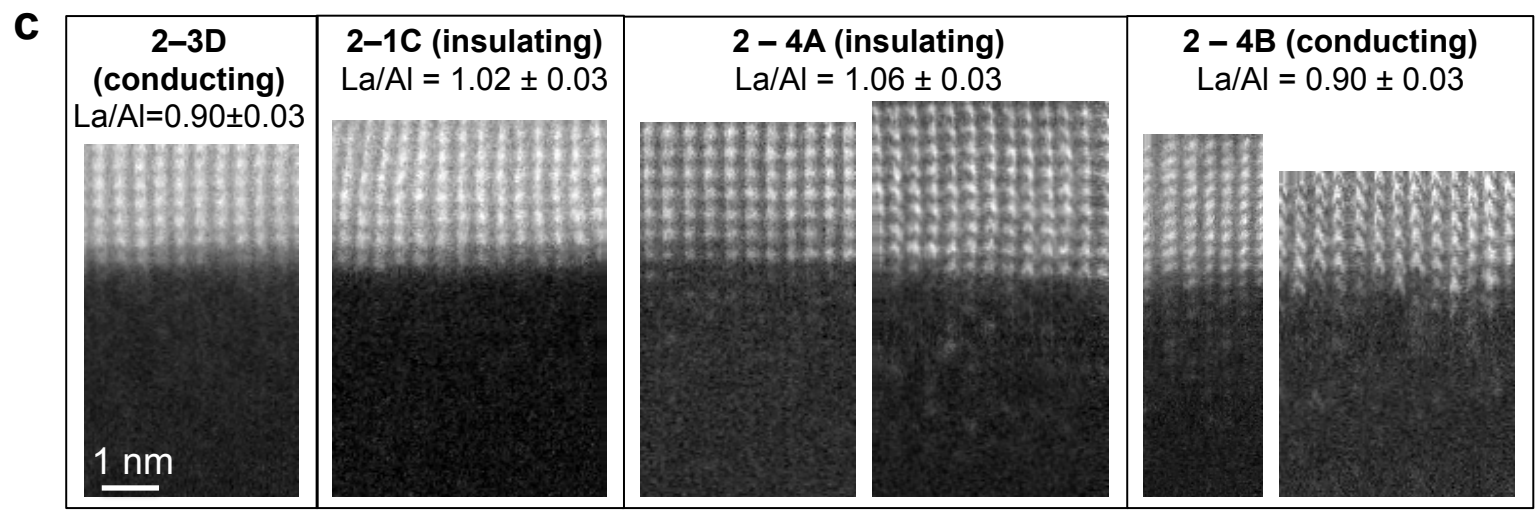

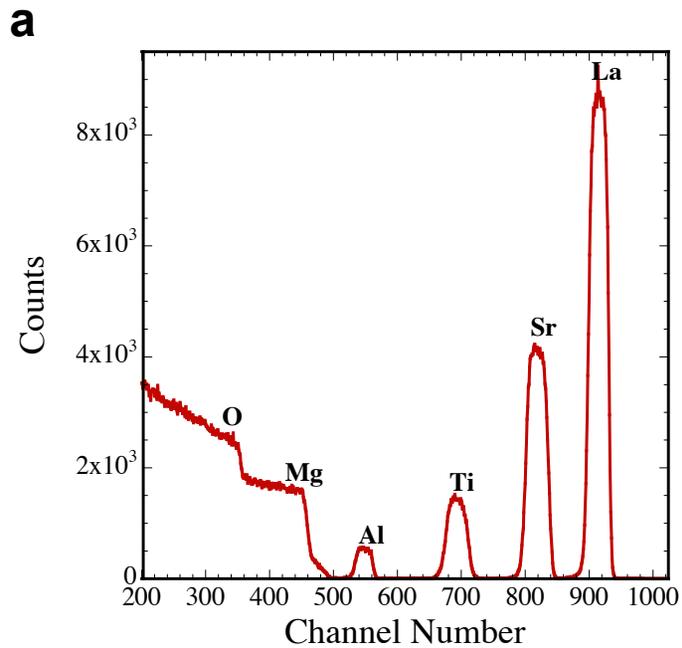

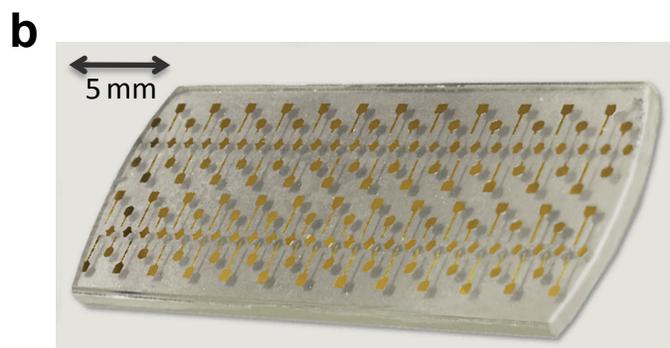

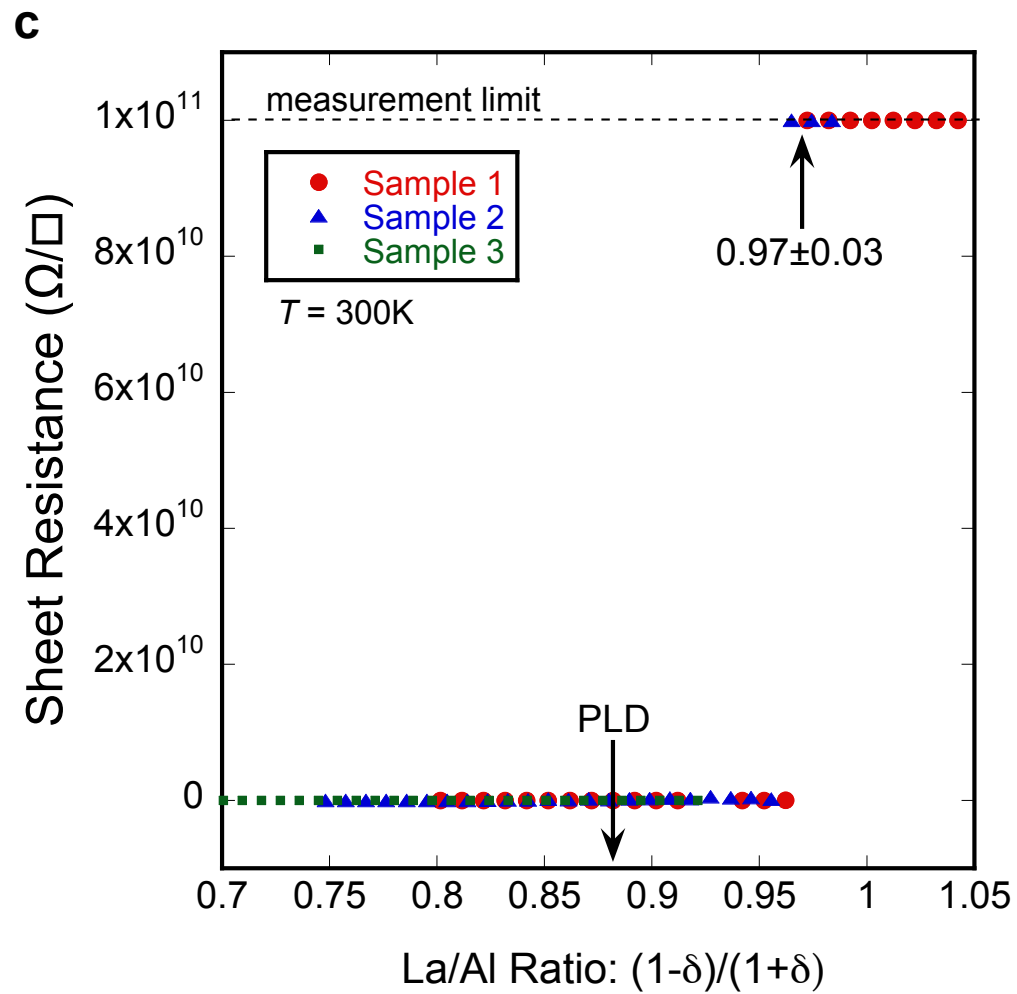

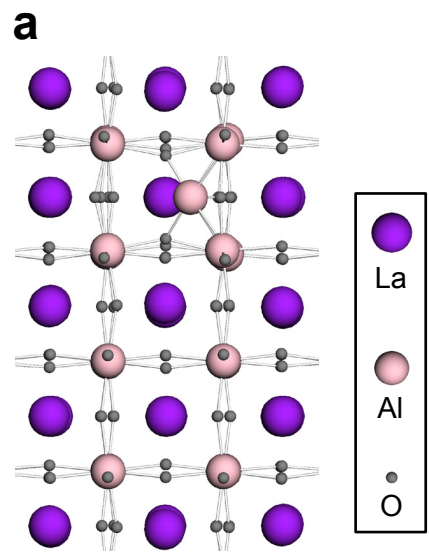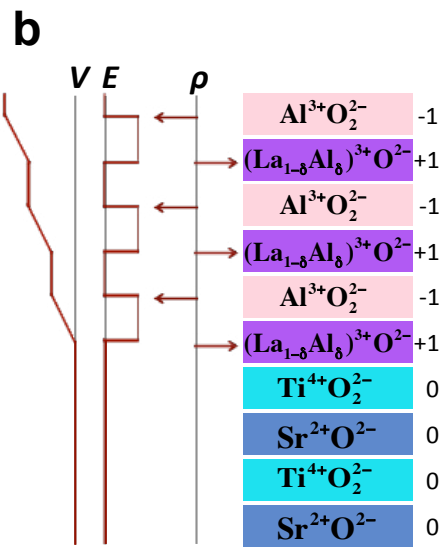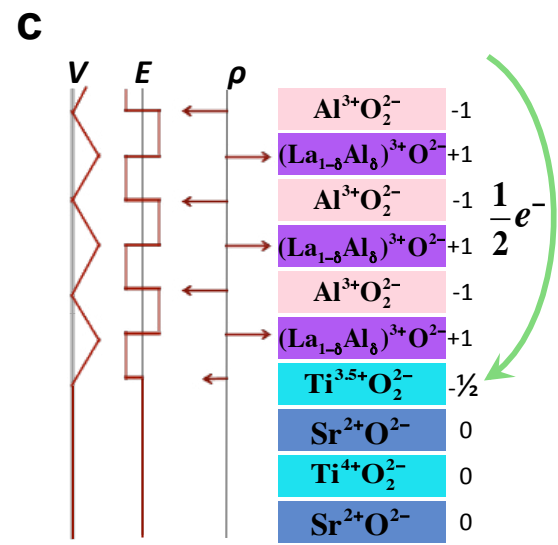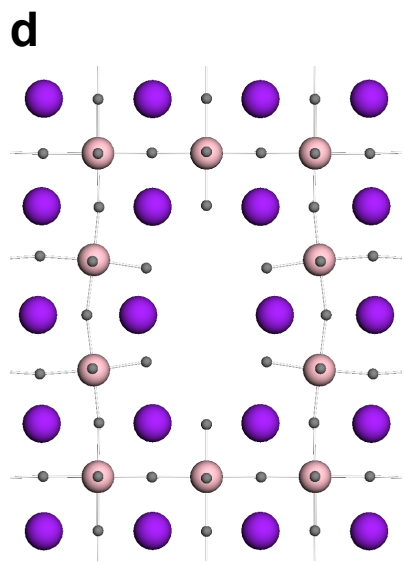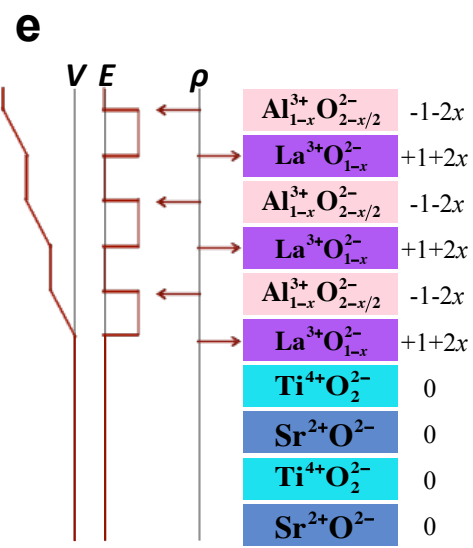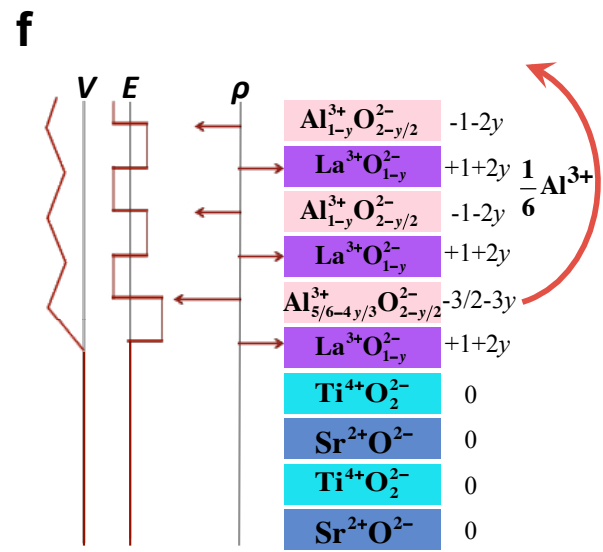

# THEORY

## a. Al-rich ($\delta = 0.05$)

| Layer | Charge |
|---|---|
| $Al^{3+}O_2^{2-}$ | −1 |
| $(La_{1-\delta}Al_\delta)^{3+}O^{2-}$ | +1 |
| $Al^{3+}O_2^{2-}$ | −1 |
| $(La_{1-\delta}Al_\delta)^{3+}O^{2-}$ | +1 |
| $Al^{3+}O_2^{2-}$ | −1 |
| $(La_{1-\delta}Al_\delta)^{3+}O^{2-}$ | +1 |
| $Ti^{3.5+}O_2^{2-}$ | −½ |
| $Sr^{2+}O^{2-}$ | 0 |
| $Ti^{4+}O_2^{2-}$ | 0 |
| $Sr^{2+}O^{2-}$ | 0 |

$\frac{1}{2}e^-$

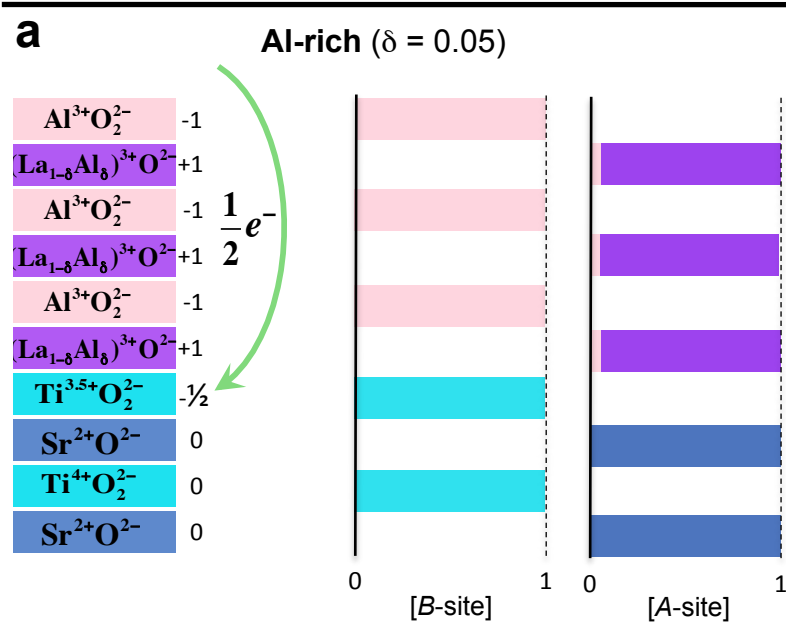

## c. La-rich ($y = 0.04$)

| Layer | Charge |
|---|---|
| $Al^{3+}_{1-y}O^{2-}_{2-y/2}$ | $-1-2y$ |
| $La^{3+}O^{2-}_{1-y}$ | $+1+2y$ |
| $Al^{3+}_{1-y}O^{2-}_{2-y/2}$ | $-1-2y$ |
| $La^{3+}O^{2-}_{1-y}$ | $+1+2y$ |
| $Al^{3+}_{5/6-4y/3}O^{2-}_{2-y/2}$ | $-3/2-3y$ |
| $La^{3+}O^{2-}_{1-y}$ | $+1+2y$ |
| $Ti^{4+}O_2^{2-}$ | 0 |
| $Sr^{2+}O^{2-}$ | 0 |
| $Ti^{4+}O_2^{2-}$ | 0 |
| $Sr^{2+}O^{2-}$ | 0 |

$\frac{1}{6}Al^{3+}$

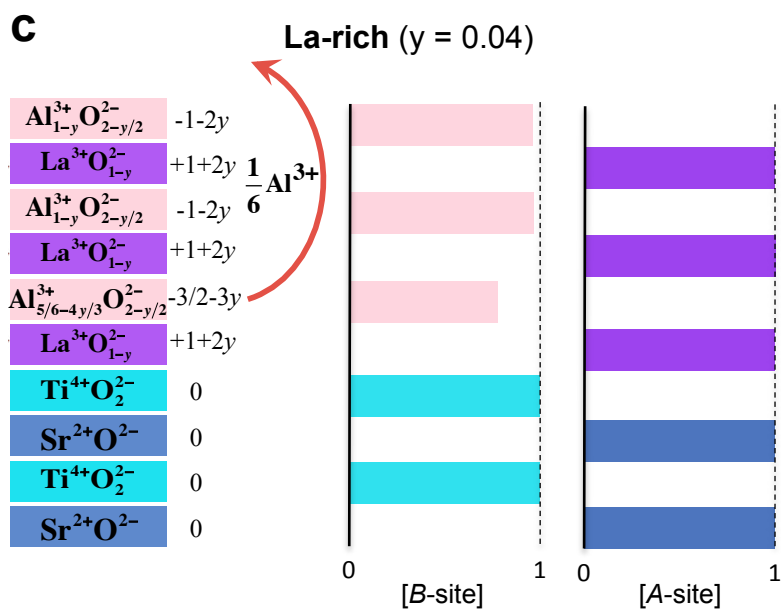

# EXPERIMENT

## b. 2 – 4B (conducting) La/Al = 0.90 ± 0.03

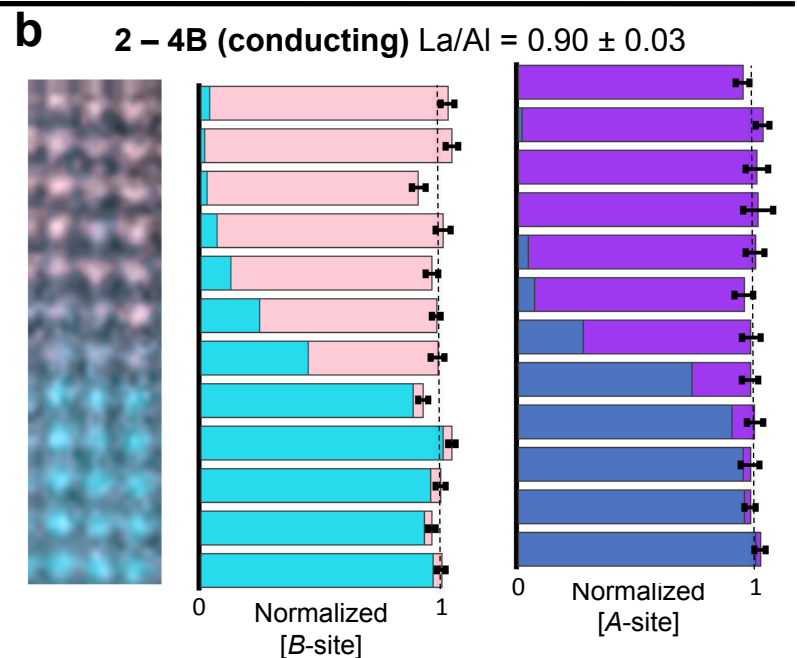

## d. 2 – 4A (insulating) La/Al = 1.06 ± 0.03

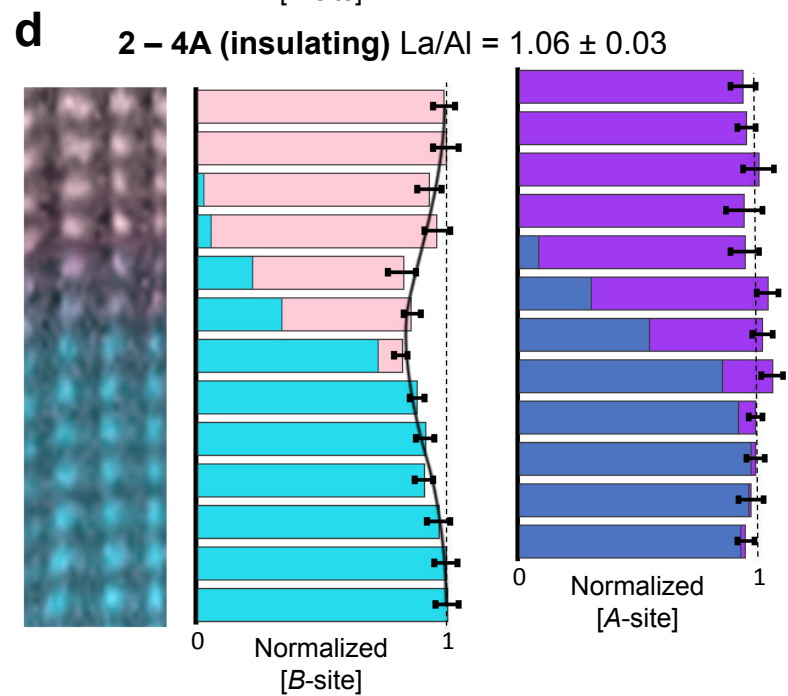

**Supplementary Information**

**LaAlO$_3$ stoichiometry found key to electron liquid formation at LaAlO$_3$/SrTiO$_3$ interfaces**

M. P. Warusawithana[1*], C. Richter[2,3], J. A. Mundy[4], P. Roy[1], J. Ludwig[1], S. Paetel[2], T. Heeg[5], A. A. Pawlicki[1], L. F. Kourkoutis[4,6], M. Zheng[7], M. Lee[1], B. Mulcahy[7], W. Zander[8], Y. Zhu[4], J. Schubert[8], J. N. Eckstein[7], D. A. Muller[4,6], C. Stephen Hellberg[9], J. Mannhart[3], D. G. Schlom[5,6]

[1] National High Magnetic Field Laboratory, Department of Physics, Florida State University, Tallahassee, Florida 32310, USA
[2] Center for Electronic Correlations and Magnetism, University of Augsburg, D-86135 Augsburg, Germany
[3] Max-Planck-Institut für Festkörperforschung, Heisenbergstrasse 1, D-70569 Stuttgart, Germany
[4] School of Applied and Engineering Physics, Cornell University, Ithaca, New York 14853, USA
[5] Department of Materials Science and Engineering, Cornell University, Ithaca, New York 14853, USA
[6] Kavli Institute at Cornell for Nanoscale Science, Ithaca, New York 14853, USA
[7] Department of Physics, University of Illinois at Urbana-Champaign, Urbana, Illinois 61801, USA
[8] Peter Gruenberg Institute 9, JARA-Fundamentals of Future Information Technologies, Research Centre Jülich, D-52425 Jülich, Germany
[9] Center for Computational Materials Science, Naval Research Laboratory, Washington, DC 20375, USA

*e-mail: maitri@magnet.fsu.edu

<u>**Supplementary Discussion 1.**</u>
**Lanthanum and aluminum flux calibration by homoepitaxial growth**

The precise calibration of lanthanum and aluminum fluxes prior to the growth of $La_{(1-\delta)}Al_{(1+\delta)}O_3$ on (100) $SrTiO_3$ substrates is critical to the results discussed in this study. To achieve this calibration, the temperatures of both the lanthanum and the aluminum sources were first adjusted to obtain a flux of approximately $1.3 \times 10^{13}$ atoms/(cm$^2 \cdot$s) for each element. The approximate fluxes were measured with a quartz crystal microbalance (QCM) using tooling factors predetermined for each source by *ex situ* Rutherford back scattering measurements. With the lanthanum and aluminum fluxes approximately matched, homoepitaxial $LaAlO_3$ was grown on (100) $LaAlO_3$ substrates via codeposition (both lanthanum and aluminum shutters were simultaneously opened and left open) in a distilled ozone background partial pressure of $1 \times 10^{-6}$ Torr and a substrate temperature of ~680 ºC. The strategy here was to exactly match the lanthanum and aluminum fluxes by carefully adjusting the aluminum source temperature using *in situ* reflection high-energy electron diffraction (RHEED) feedback obtained during this homoepitaxial $LaAlO_3$ growth. This process is discussed below.

The calibration sample was continuously rotated during growth. This allowed for simultaneous monitoring of the RHEED patterns of the growing surface layer along multiple in-plane crystallographic directions. Weak, but distinct signatures in the RHEED patterns were empirically determined that corresponded to slight excesses of lanthanum or aluminum in the surface layer. Fig. S1 shows a representative set of RHEED images obtained at the end of a homoepitaxial calibration growth once the lanthanum and aluminum fluxes were perfectly matched. The distilled ozone partial pressure was maintained at $1 \times 10^{-6}$ Torr as these images were acquired. The top row of images, Fig. S1a to Fig. S1c, show RHEED patterns obtained along the [100], [110] and [210] azimuths, respectively, from the neutral surface with no excess of either lanthanum or aluminum atoms. Sharp RHEED patterns are observed along all three directions just like that of a $LaAlO_3$ substrate with no extra streaks characteristic of surface reconstructions. Subsequently, the center row of images, Fig. S1d to Fig. S1f, were obtained after depositing ~7% excess of aluminum achieved by opening only the aluminum shutter for just 4 seconds (~55 seconds corresponds to a monolayer of aluminum). Weak half-order streaks can be observed along both the [100] azimuth (Fig. S1d) and the [210] azimuth (Fig. S1f) and are indicated by yellow arrows. Finally, the bottom row of images, Fig. S1g to Fig. S1i, were captured from a surface with ~7% excess of lanthanum achieved by depositing only lanthanum for ~8 seconds. Note that the first 4 seconds of lanthanum balanced the excess aluminum on the surface leading to neutral RHEED patterns (as shown in Fig. S1a to Fig. S1c) while the last 4 seconds led to ~7% excess of lanthanum. The lanthanum rich surface is characterized by half order streaks only along the [110] azimuth indicated by yellow arrows (Fig. S1h). Note that if the surface becomes much more lanthanum rich, faint half-order streaks could appear along the [210] azimuth as well (not shown), but not along the [100] azimuth.

The half-order streaks that appear in the RHEED patterns during growth of the homoepitaxial $LaAlO_3$ calibration film along the different in-plane directions provided sensitive feedback enabling the lanthanum and aluminum fluxes to be matched. If the RHEED patterns began to show signatures of an aluminum (lanthanum) rich surface, as lanthanum and aluminum are codeposited starting from a neutral surface, first, the aluminum (lanthanum) shutter was closed for a few seconds at a time to reduce the aluminum (lanthanum) content of the surface layer until the surface was reversed. i.e., RHEED showed signatures of a surface rich in the other species.

At this point the aluminum source temperature was lowered (increased) on the fly by as small a step as 0.1 ºC, which corresponds to a change in aluminum flux as low as ~0.2%. With this change in aluminum flux, the RHEED patterns were closely monitored. If after some time of codeposition, the RHEED patterns still developed characteristic features corresponding to the excess of one species, the above steps were repeated to reduce the difference in the flux between lanthanum and aluminum. This process was carried out until the LaAlO₃ calibration sample could be grown via codeposition for over 20 minutes (~20 unit cells thick) with the RHEED indicating a neutral surface layer without having to close the shutters or make any temperature adjustments to the sources. Based on the assumption that an excess of a species accumulates mostly at the surface, at this point the lanthanum and aluminum fluxes are matched to better than 0.4% since this would result in over a 7% excess at the surface during the 20 layers of growth and would be easily observed in RHEED as shown in Fig. S1. With the lanthanum and aluminum fluxes matched to better than 0.4%, the calibration growth was terminated and the (001) SrTiO₃ substrate was loaded into the chamber and the growth of the actual sample immediately followed.

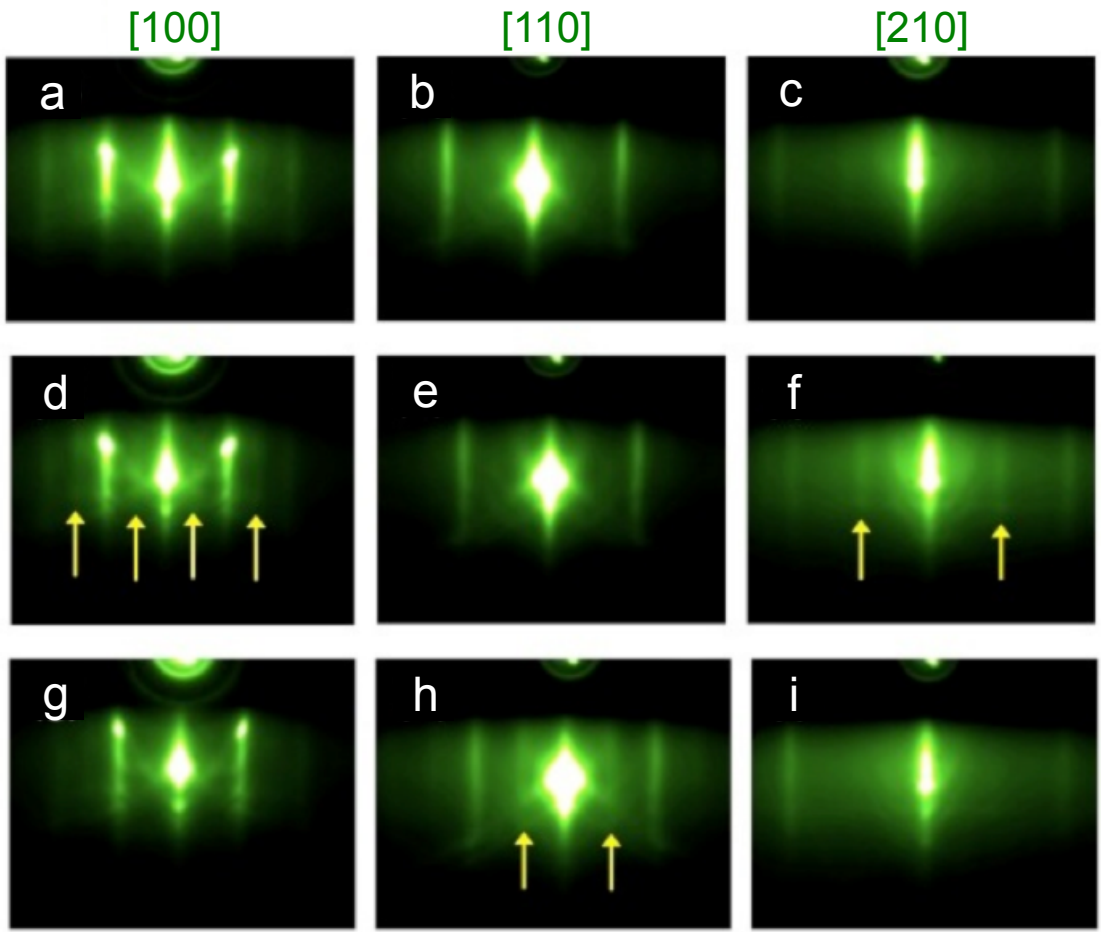

**Figure S1. Sensitivity of surface reconstruction to $La_{(1-\delta)}Al_{(1+\delta)}O_3$ stoichiometry. a-c,** RHEED images taken of the neutral surface of a homoepitaxial LaAlO₃ film. No half-order streaks were seen along [100], [110] and [210] azimuths. **d-f,** RHEED images taken after opening the aluminum shutter for 4 s. This made the surface aluminum rich. Half-order maxima were seen along [100] and [210] azimuths as indicated by the arrows. **g-i,** RHEED images taken after making the surface slightly lanthanum rich. The arrows indicate half-order streaks seen along the [110] azimuth due to a slight excess of lanthanum.

<u>**Supplementary Discussion 2.**</u>
**Cation stoichiometry via Rutherford backscattering spectrometry (RBS) measurements**

The accuracy with which the cation stoichiometry can be predicted relies on the accuracy with which the RBS data can be extracted and how well the different peaks can be separated in the RBS spectrum. With aluminum being a light element, extracting accurate cation stoichiometry using RBS measurements for a film containing aluminum is inherently a very challenging process. This is because peaks of constituent elements of many substrates (e.g., $SrTiO_3$) would bury the aluminum peak. Even if the $LaAlO_3$ film were grown on a silicon substrate for RBS measurements, the aluminum peak would be underneath the silicon peak making it impossible to extract an accurate number density for aluminum in the sample. Similarly, if an MgO substrate were used for this purpose, the aluminum peak would be riding the shoulder of the magnesium peak leading to the same problem. Thus an important part of this experiment was the specially structured companion sample that was used for all RBS measurements. The companion sample was grown on a (100) MgO substrate on which an approximately 500Å thick buffer layer of $SrTiO_3$ was grown prior to growing the $LaAlO_3$ layer that was to be calibrated. The $LaAlO_3$ layer was also approximately 400 Å thick in order to obtain a low statistical error and was grown immediately following the device sample. The purpose of the intermediate $SrTiO_3$ film was to lower the energy of the alpha particles as they penetrated through the intermediate film before being scattered off the substrate. This moved the substrate magnesium peak to sufficiently lower energies such that the aluminum peak was clearly separated from the magnesium peak. From the RBS spectrum of the companion sample (Fig. 3a), by peak integration, an accurate measurement of the La/Al ratio at the center of the device sample was determined.

To obtain the spatial variation of the La/Al ratio across the $La_{(1-\delta)}Al_{(1+\delta)}O_3$ films grown on the long (100) $SrTiO_3$ crystals, two separate RBS calibration samples were grown, where the calibration samples also spanned the spatial extent of the long $SrTiO_3$ samples. These RBS calibration samples were structured similarly to the companion RBS samples. RBS data was measured as a function of position on each of these two calibration growths. It should be noted that the spatial variation of the La/Al ratio, $(1-\delta)/(1+\delta)$, is dependent only on the configuration of the MBE system when the same growth conditions are used. From the spatially resolved RBS measurements it was observed that, for our system, the variation in the La/Al ratio was mostly along the left-to-right ($x$) direction. To obtain the *relative spatial variation* of the La/Al ratio, $(1-\delta)/(1+\delta)$, along the $x$-direction for any sample, the La/Al ratio at the center of each of the calibration RBS samples was scaled to $(1-\delta)/(1+\delta) = 1.00$. This data (Fig. S2) which provides the spatial variation of the La/Al ratio across a sample along the $x$ direction, together with the data from the companion RBS sample which provides the absolute La/Al ratio at the center of the device sample, makes it possible to accurately determine the La/Al ratio, $(1-\delta)/(1+\delta)$ at any point of the long $La_{(1-\delta)}Al_{(1+\delta)}O_3$ / (100) $SrTiO_3$ samples.

<u>RBS error analysis:</u>
The sheet resistance of each device of the three long samples is plotted in Fig. 3c and in Supplementary Fig. S5 as a function of the La/Al ratio, $(1-\delta)/(1+\delta)$, obtained as discussed above. The error associated with the La/Al ratio, is due to the convolution of four sources of error:
1. RBS counting error (0.5% - 1.0%).
2. Spatial determination of the RBS measurement spot position (1.5% - 2%).
3. Drift in source fluxes between consecutive growths: from device sample to companion RBS sample (1% - 2%).
4. Device position/placement (±1 mm) from center of sample (1% - 1.2%).

The estimated range for each error source is shown in parenthesis. This leads to a convoluted error in $(1-\delta)/(1+\delta)$ of 2% (taking the lower estimates) or 3% (from the more conservative estimates). The conservative error estimate is reported in the manuscript.

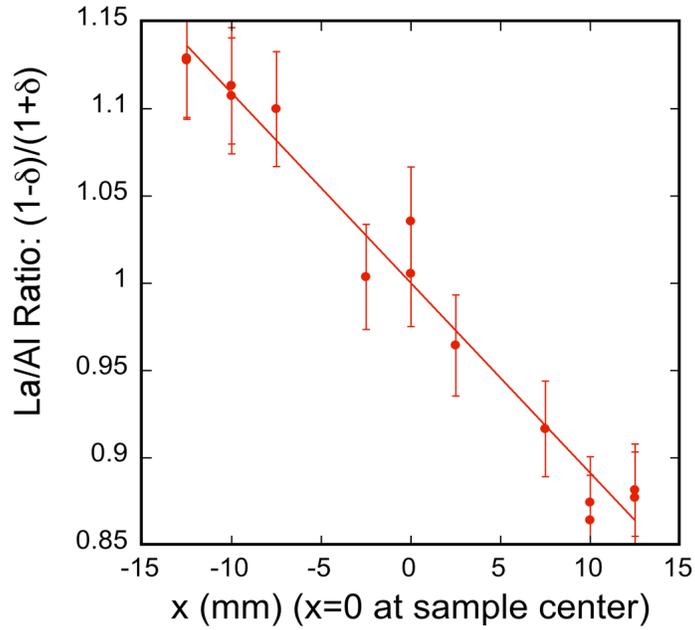

**Figure S2. Relative spatial variation of La/Al ratio along the left-to-right ($x$) direction due to the placement of sources in MBE system.** The errors involved are discussed under "RBS error analysis."

**Supplementary Figure S3.**
**EELS fine structure of the O-*K* edge of the LaAlO$_3$/SrTiO$_3$ interface for a conducting and an insulating sample**

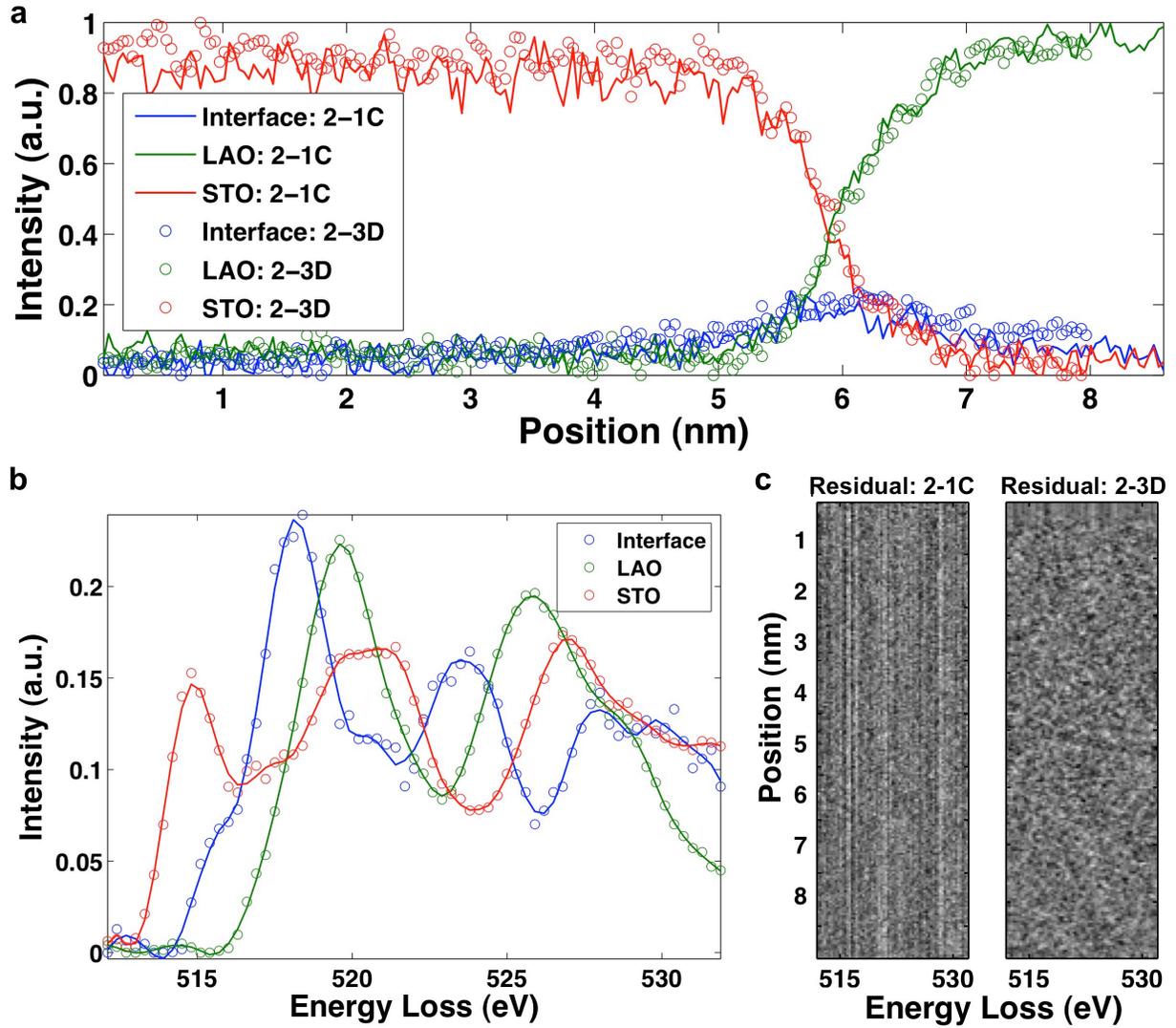

**Figure S3. EELS fine structure of the O-*K* edge of the LaAlO$_3$/SrTiO$_3$ interface for a conducting (2–3D) and an insulating (2–1C) sample. a,** The concentration profile of the data fit to a LaAlO$_3$ and SrTiO$_3$ reference spectra as well as a distinct interface component extracted from the data by multivariate statistical analysis. **b**. Spectra used in the fit. Both samples show the presence of an interfacial O-*K* edge component, distinct from LaAlO$_3$ and SrTiO$_3$. This component appears, however, for both samples with the same concentration and spatial extent. **c,** The residual from the fit showing the absence of any remaining structure. The streaks for 2–1C result from a slightly different fixed pattern noise in the spectra due to a different acquisition time.

**Supplementary Figure S4.**
**STEM-EELS fine structure of the Ti-$L_{2,3}$ edge of the LaAlO$_3$/SrTiO$_3$ interface for a conducting (2–3D) and an insulating (2–1C) sample**

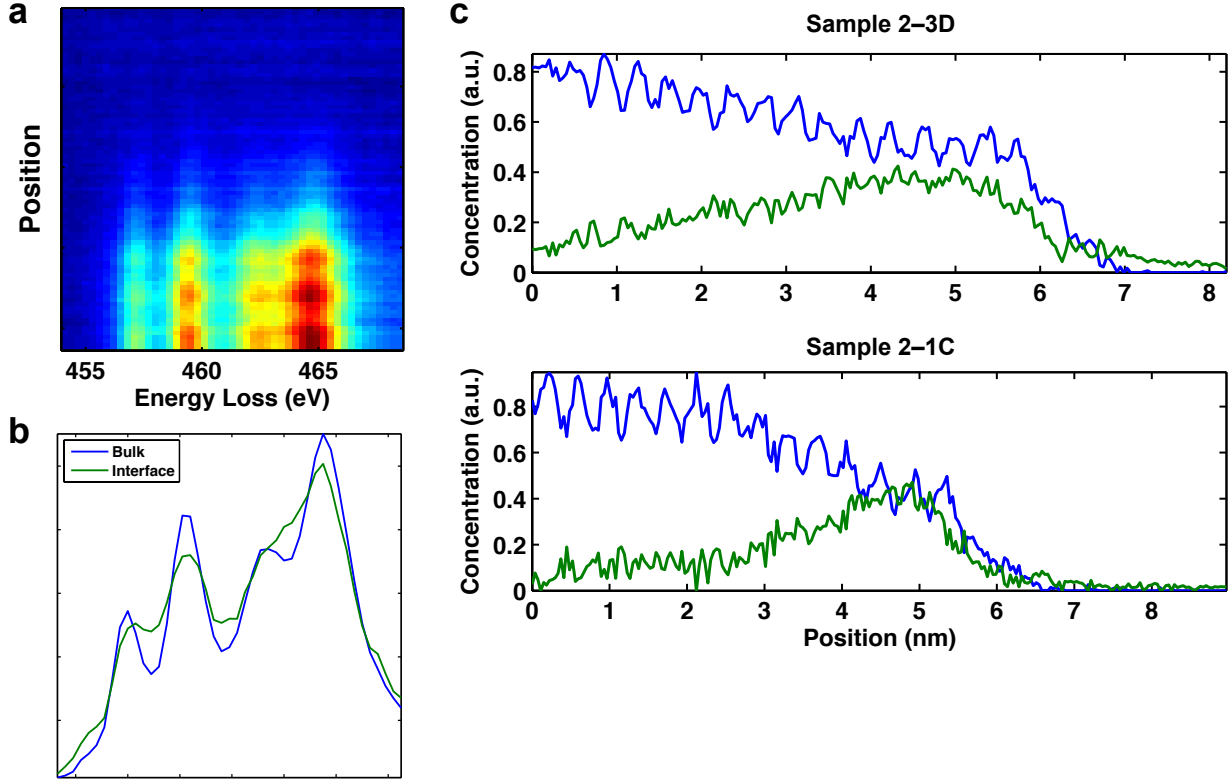

**Figure S4. STEM-EELS fine structure of the Ti-$L_{2,3}$ edge of the LaAlO$_3$/SrTiO$_3$ interface for a conducting (2–3D) and an insulating (2–1C) sample. a,** A two-dimensional profile of the Ti-$L_{2,3}$ edge across the LaAlO$_3$/SrTiO$_3$ interface for the conducting interface 2–3D. The interface is interdiffused, although no shift to lower onset energy, indicative of a reduction in the titanium valence, is observed. **b**, Spectra extracted from the multivariate curve resolution analysis of the Ti-$L_{2,3}$ edge. The "interface" component tracks a reduction in the crystal field splitting from the titanium edge, possibly caused by local lattice distortions near the interface to SrTiO$_3$. The Ti-$L_{2,3}$ edge spectrum at the interface shows a significant component that is different from the bulk. A large portion is from crystal field distortions, but we cannot exclude the possibility of a small (<0.1 eV) contribution from Ti$^{+3}$. The spectra and the lack of a dominant Ti$^{+3}$ component look similar to those reported for other high oxygen pressure growths of LaAlO$_3$/SrTiO$_3$[1,2]. **c,** Result of the fit of the spectra in (b) to the full data sets. Deviations in the quantity of the "interface" component observed could be due to slight differences in lanthanum interdiffusion for the two samples.

**Supplementary Figure S5.**
**Room temperature sheet resistance of $La_{(1-\delta)}Al_{(1+\delta)}O_3$ / (100) $SrTiO_3$ interfaces as a function of La/Al ratio**

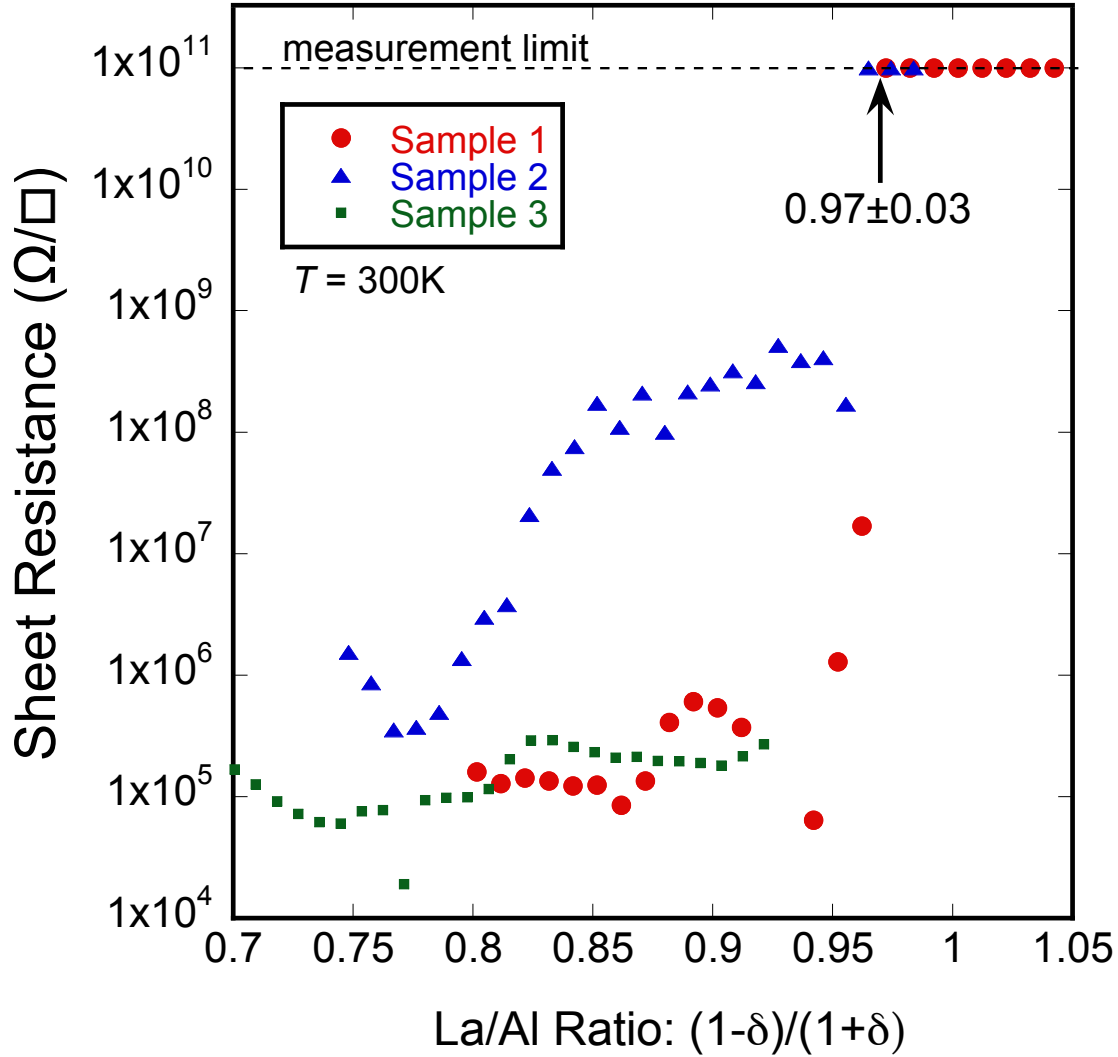

**Figure S5. Room temperature sheet resistance of $La_{(1-\delta)}Al_{(1+\delta)}O_3$ / (100) $SrTiO_3$ interfaces as a function of La/Al ratio.** The data shown in Fig. 3c is plotted here on a logarithmic resistance scale. The La/Al ratio was determined by RBS measurements. Local 4-point resistance measurements were made on all devices placed at 1 mm intervals on three long samples. A sharp jump in sheet resistance is observed at a La/Al ratio of 0.97±0.03, consistent in all three samples. See Supplementary Discussion 2 for error analysis. The overall higher interface resistance observed in sample 2 is likely due to slight variations in substrate/substrate surface termination. Also, on two consecutive devices of sample 1, accurate measurements were not possible presumably due to sample processing issues. These two data points have been removed from the plot.

## Supplementary Discussion 3.
## First principles predictions of defect structure and electronic properties of off-stoichiometric La$_{(1-\delta)}$Al$_{(1+\delta)}$O$_3$ films

The density functional calculations used the generalized-gradient approximation[3] and projector-augmented wave functions[4] as implemented in VASP[5,6]. A 282.8 eV planewave cutoff was used for the basis set. In all calculations, the in-plane lattice constant was fixed to the theoretical SrTiO$_3$ lattice constant, $a_{SrTiO_3}$ = 3.948 Å.

The section can be outlined as follows:
- Defects form when LaAlO$_3$ is grown off stoichiometry.
- The extended Al-O vacancy complexes formed in lanthanum-rich films allow ionic motion between the interface and surface while the localized aluminum-rich defects do not.
- In the limit of thick LaAlO$_3$ films, cation vacancies will form spontaneously at the interface to compensate the diverging potential[7] if given the chance, but this mechanism is negligible in thin films.
- In lanthanum-rich films, the extended defects in the films allow cation vacancies to move to the interface to compensate the diverging potential in thicker films. The localized defects in the aluminum-rich films allow negligible cation motion, and a metallic interface forms after the film reaches a critical thickness.

First we determine which defects form in bulk La$_{(1-\delta)}$Al$_{(1+\delta)}$O$_3$ strained to the SrTiO$_3$ lattice constant. La/Al stoichiometries of approximately $(1-\delta)/(1+\delta)$ = 15/16 and 16/15 (0.9375 and 1.0667) were used. For the aluminum-rich calculations, the Al$_{La}$ substitutional defect shown in the main paper had an energy more than 60 meV per unit cell lower than the lowest energy La-O vacancy complex that we found[8].

In the lanthanum-rich case, the lowest energy Al-O vacancy complexes are columns of (Al$_2$O$_3$)$_n$ vacancies, shown in Fig. S6. As the width $n$ of the vacancy columns increases, the energy decreases roughly linearly with $1/n$ as seen in Fig. S7. For $n \geq 8$, the energy of the (Al$_2$O$_3$)$_n$ vacancy column is less than the energy of the La$_{Al}$ substitutional defect, shown in Fig. S8. Even for $n < 8$, once an (Al$_2$O$_3$)$_n$ vacancy forms in one layer, it is likely to persist in subsequently grown layers.

To compare structures with different stoichiometries energetically, we compute the Gibbs free energy,

$$G = E - \mu_{La} N_{La} - \mu_{Al} N_{Al} - \mu_O N_O, \quad (1)$$

where $E$ is the total energy of the system computed with DFT, the $N_i$ are the numbers of each atom in the calculation, and the $\mu_i$ are the chemical potentials. The potentials are subject to the constraint

$$\mu_{La} + \mu_{Al} + 3\mu_O = \mu_{LaAlO_3}^{bulk}, \quad (2)$$

where $\mu_{LaAlO_3}^{bulk}$ is the computed energy of bulk LaAlO$_3$ strained in the $x$-$y$ plane to the SrTiO$_3$ lattice constant and relaxed in the $z$ direction. The growth conditions (680 °C and 10$^{-6}$ Torr of distilled ozone) determine $\mu_O$[9]. Thus there is only one independent parameter: We will vary $\mu_{La}$ and determine $\mu_{Al}$ from Eq. (2).

The energies of the stoichiometric, aluminum-rich, and lanthanum-rich phases of bulk $La_{(1-\delta)}Al_{(1+\delta)}O_3$ strained to the $SrTiO_3$ lattice constant are plotted in Fig. S9. We use this plot to determine the ranges of $\mu_{La}$ over which each phase is stable. For $\mu_{La} < -6.73$ eV, the film is aluminum-rich, while for $\mu_{La} > -5.91$ eV, it is lanthanum-rich.

Structures with higher energies are not shown. There has been interest in the role of oxygen vacancies in this system. We find that the formation energy of oxygen vacancies in bulk $LaAlO_3$ is high and is not affected by either $Al_2O_3$ vacancy complexes or $La_{Al}$ substitutional defects – The oxygen vacancy formation energies are the same as in stoichiometric $LaAlO_3$[10].

We now turn to calculations of $La_{(1-\delta)}Al_{(1+\delta)}O_3$ films on $SrTiO_3$. In the limit of thick ideal $LaAlO_3$ films on $SrTiO_3$, ½ electron is transferred to the $SrTiO_3$ conduction band, and there is no electric field in the $LaAlO_3$[7,11]. The surface reconstructs to assume a positive charge; it has been shown that oxygen vacancies form spontaneously at the surface of sufficiently thick $LaAlO_3$ on $SrTiO_3$[12]. Note that oxygen vacancies in the $SrTiO_3$ substrate, however, do not screen the field in the $LaAlO_3$ since these vacancies lie on the same side of the $LaAlO_3$ as the electrons in the $SrTiO_3$ conduction band. We approximate the limit of thick $LaAlO_3$ films using the multilayer structure shown in Fig. S10, with two identical interfaces and no surfaces. The $LaAlO_3$ has one more LaO layer than $AlO_2$ layers, resulting in 1 electron in the conduction band, which is shared by the two interfaces. Due to mirror symmetry, there is no electric field in the middle of the $LaAlO_3$.

We examine the stability of charged defects at the $La_{(1-\delta)}Al_{(1+\delta)}O_3/SrTiO_3$ interface by putting lanthanum and aluminum vacancies in the multilayer calculations, keeping the mirror symmetry. The vacancy structures are shown in Fig. S11. At a vacancy density of 1/6 monolayers, the +3 defects remove the 1/2 electron at the interface forming an insulator. As seen in Fig. S12, the ideal interface with the 2-DEL has higher energy than the insulating interfaces with cation vacancies under all conditions.

Interestingly, the lanthanum vacancies at the interface gain energy by diffusing into the $SrTiO_3$, forming a strontium substitutional defect and a strontium vacancy, which can diffuse far from the substitutional[13]. Thus any *A*-site deficiency will be spread out over many layers.

In summary, we have computed the nature of the defects that form in off-stoichiometric $La_{(1-\delta)}Al_{(1+\delta)}O_3$ strained to the $SrTiO_3$ lattice constant. In aluminum-rich samples, aluminum substitutes for lanthanum. In lanthanum-rich samples, $Al_2O_3$-defect-complexes form. The energetics of the defects allows us to determine the ranges of chemical potentials for each stoichiometry.

Importantly, the $Al_2O_3$-vacancy-complexes allow ionic movement between the surface and the interface, while the $Al_{La}$ substitutional defects do not. In calculations simulating thick $LaAlO_3$ films on $SrTiO_3$, we find cation vacancies form spontaneously to compensate the electron liquid.

The scenario is outlined in Fig. 4: The $La_{(1-\delta)}Al_{(1+\delta)}O_3$ films grow with defects to accommodate the stoichiometry, and the electrostatic potential difference between the surface and interfaces grows with the film thickness[7]. In the aluminum-rich films, the diverging potential is eventually screened by the formation of an interfacial electron liquid. In the lanthanum-rich films, aluminum vacancies can easily move to the interface through the $Al_2O_3$-vacancy-complexes to screen the diverging potential, and the system remains insulating.

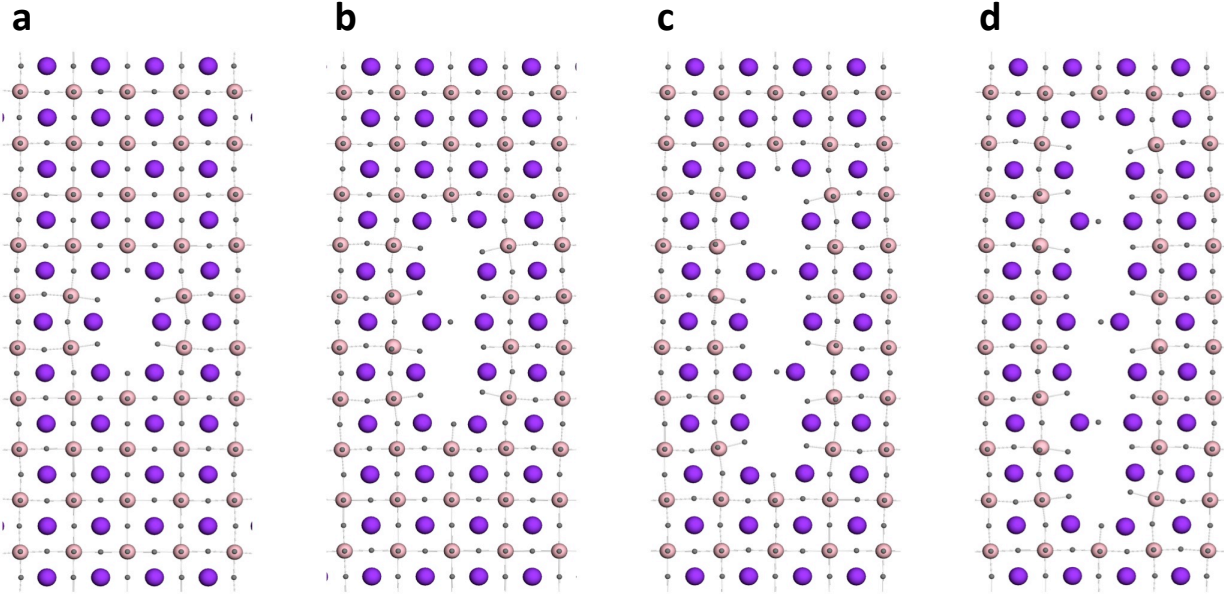

**Figure S6. Al$_2$O$_3$ vacancy complexes.** The lowest energy Al$_2$O$_3$-vacancy-complexes with widths $n$ = 2, 4, 6, and 8 are shown from above. **a,** $n$ = 2. **b,** $n$ = 4. **c,** $n$ = 6. **d,** $n$ = 8. For $n \geq 4$, there are oxygen ions in the vacancy columns. Both the oxygen ions and the nearest lanthanum ions are shifted by half a lattice constant in the $z$ direction (into the page). The stoichiometry is held fixed at approximately $(1-\delta)/(1+\delta) = 16/15$ by increasing the supercell size with increasing $n$.

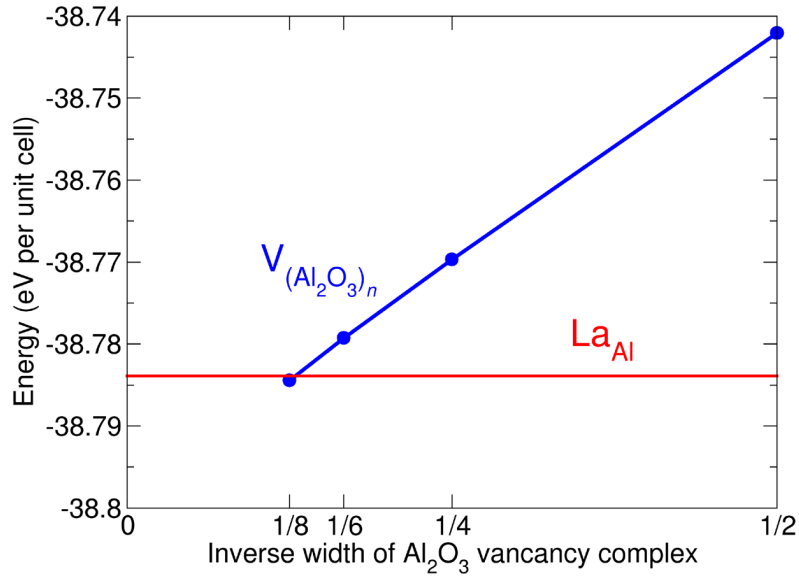

**Figure S7. Energy of the Al$_2$O$_3$-vacancy-complexes as a function of inverse width $n$.** The energy of the La$_{Al}$ substitutional defect (see Fig. S8) is shown in red.

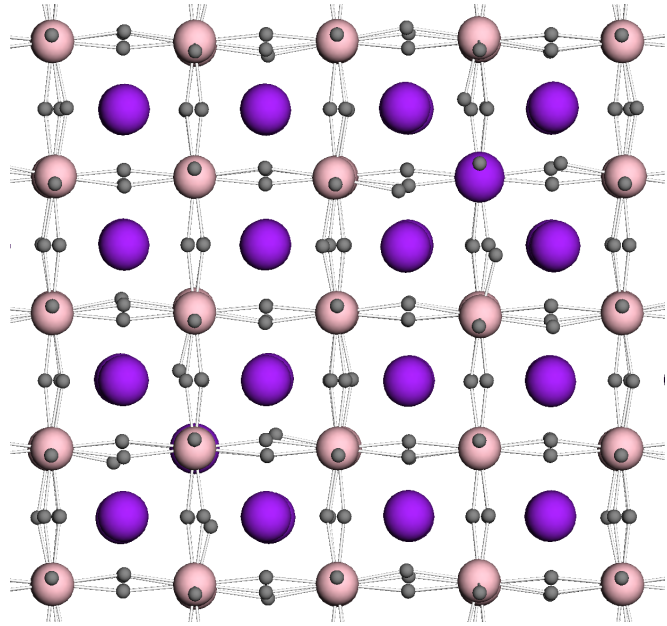

**Figure S8. La$_{Al}$ substitutional defect.** Structure of the La$_{33}$Al$_{31}$O$_{96}$ calculation with La$_{Al}$ substitutional defects viewed from the side. The stoichiometry is $(1-\delta)/(1+\delta) = 33/31$.

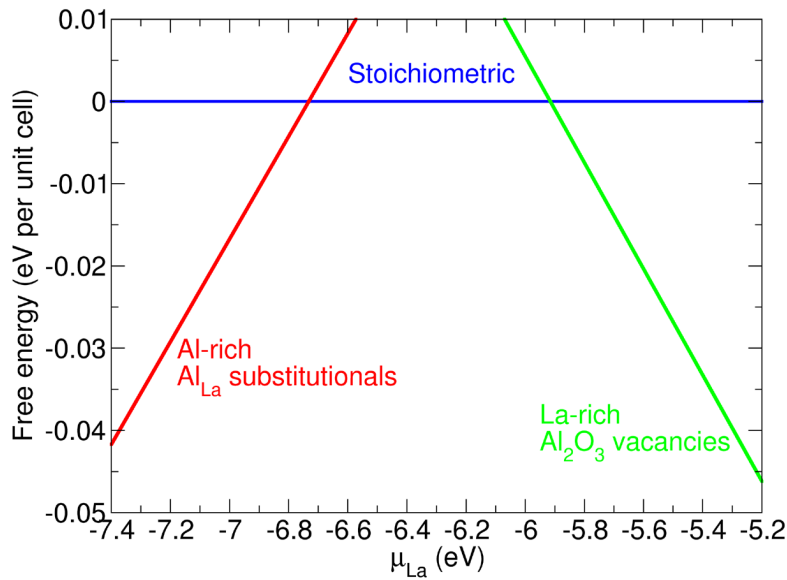

**Figure S9. Free energies of bulk La$_{(1-\delta)}$Al$_{(1+\delta)}$O$_3$ strained to the SrTiO$_3$ lattice constant at various stoichiometries.** The three lines show the free energy of ideal stoichiometric LaAlO$_3$ (blue) and the lowest energy aluminum-rich (red) and lanthanum-rich (green) structures. The stoichiometric energy is taken as the reference. The aluminum-rich compound is La$_{31}$Al$_{33}$O$_{96}$, with aluminum substituting for lanthanum as in Fig. 4a. The lanthanum-rich compound is La$_{128}$Al$_{120}$O$_{372}$, with the (Al$_2$O$_3$)$_8$-vacancy-complex in Fig. S6d. This plot is used to set the range of $\mu_{La}$ for which each stoichiometry is stable.

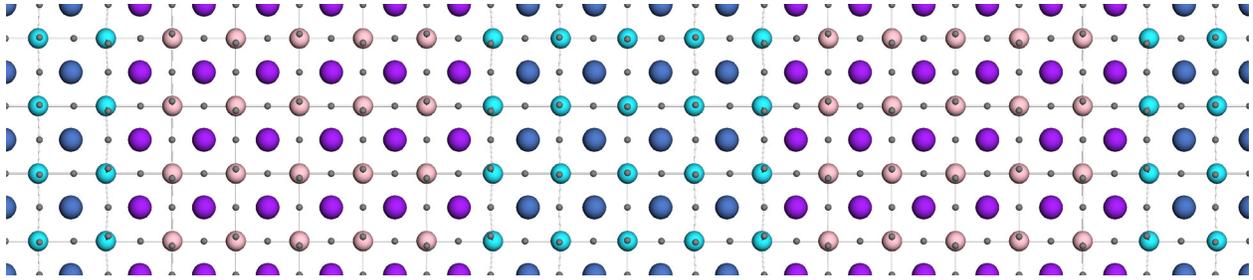

**Figure S10. Structure used for the interface computations.** A multilayer geometry consisting of 5.5 unit cells of LaAlO$_3$ (pink and purple) and 4.5 unit cells of SrTiO$_3$ (light and dark blue) was used. There are two identical defect-free LaO/TiO$_2$ interfaces, resulting in 1/2 electron per interface per areal cell in the conduction band.

**a**            **b**

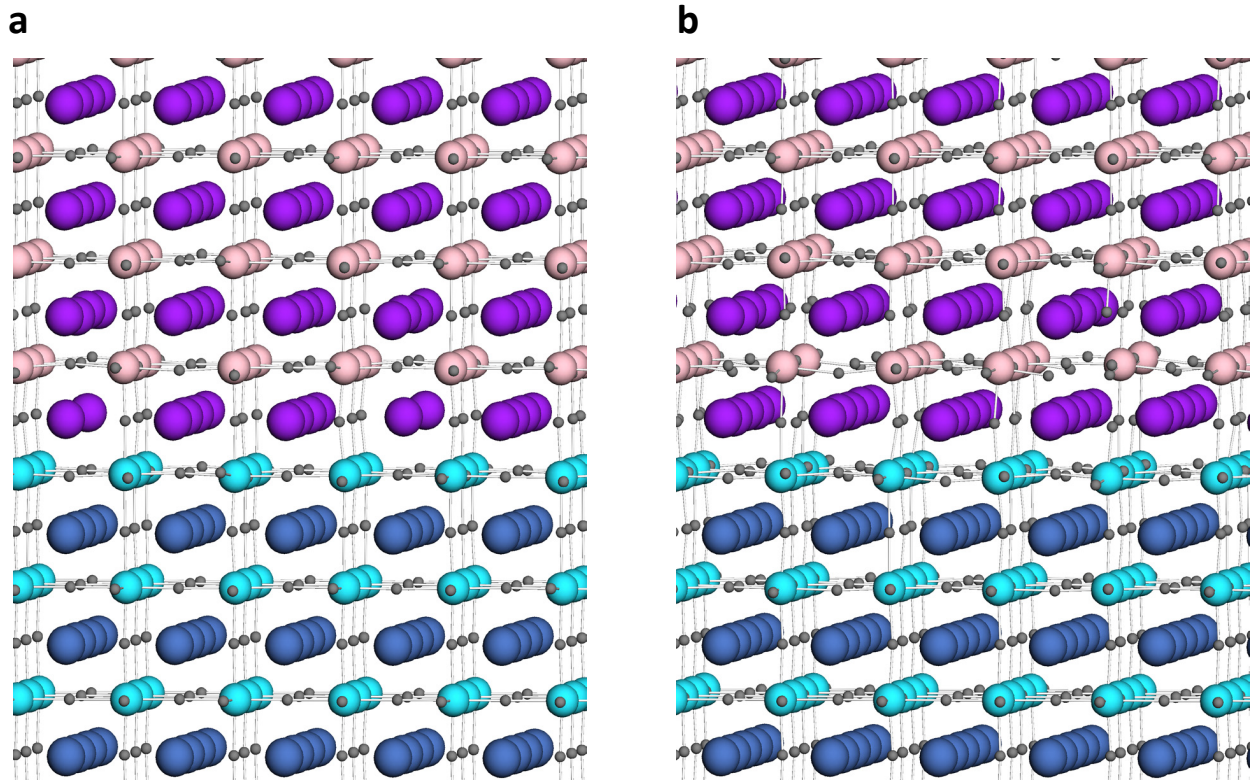

**Figure S11. The structures with cation vacancies at the interface. a,** La vacancies. **b**, Al vacancies. The vacancy densities are 1/6 monolayers. Both defects have charge +3, and the structures are insulating.

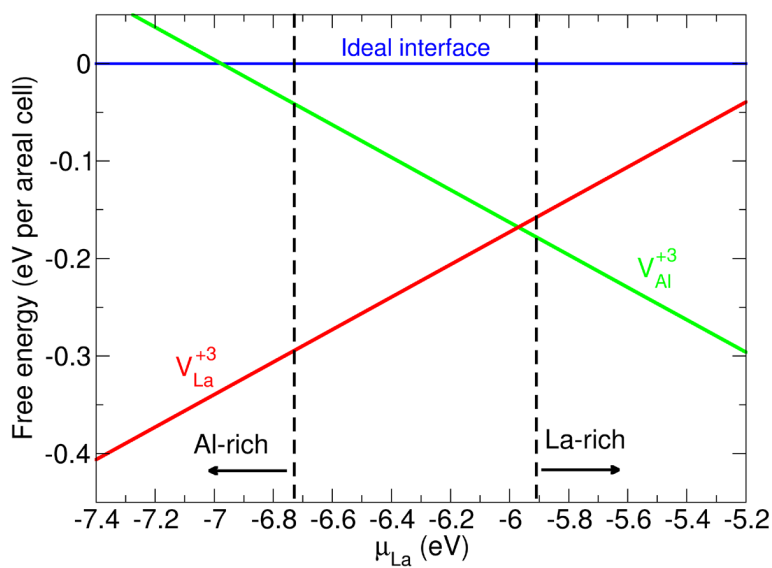

**Figure S12. Free energies of the La$_{(1-\delta)}$Al$_{(1+\delta)}$O$_3$/SrTiO$_3$ interface structures.** The ideal defect-free LaAlO$_3$/SrTiO$_3$ interface (blue line and Fig. S10) has a 2-DEL at the interface. Lanthanum vacancies with charge +3 (red line and Fig. S11a) at the interface are stable for lower $\mu_{La}$, and aluminum vacancies also charged +3 (green line and Fig. S11b) are stable for higher $\mu_{La}$. The dashed lines show the ranges of $\mu_{La}$ for each stoichiometry determined in Fig. S9. As described in the text, the aluminum-rich films prevent cation migration, and vacancies cannot form at the interface.

**Supplementary Figure S13.**
**Different views of lowest energy structures determined with density functional theory**

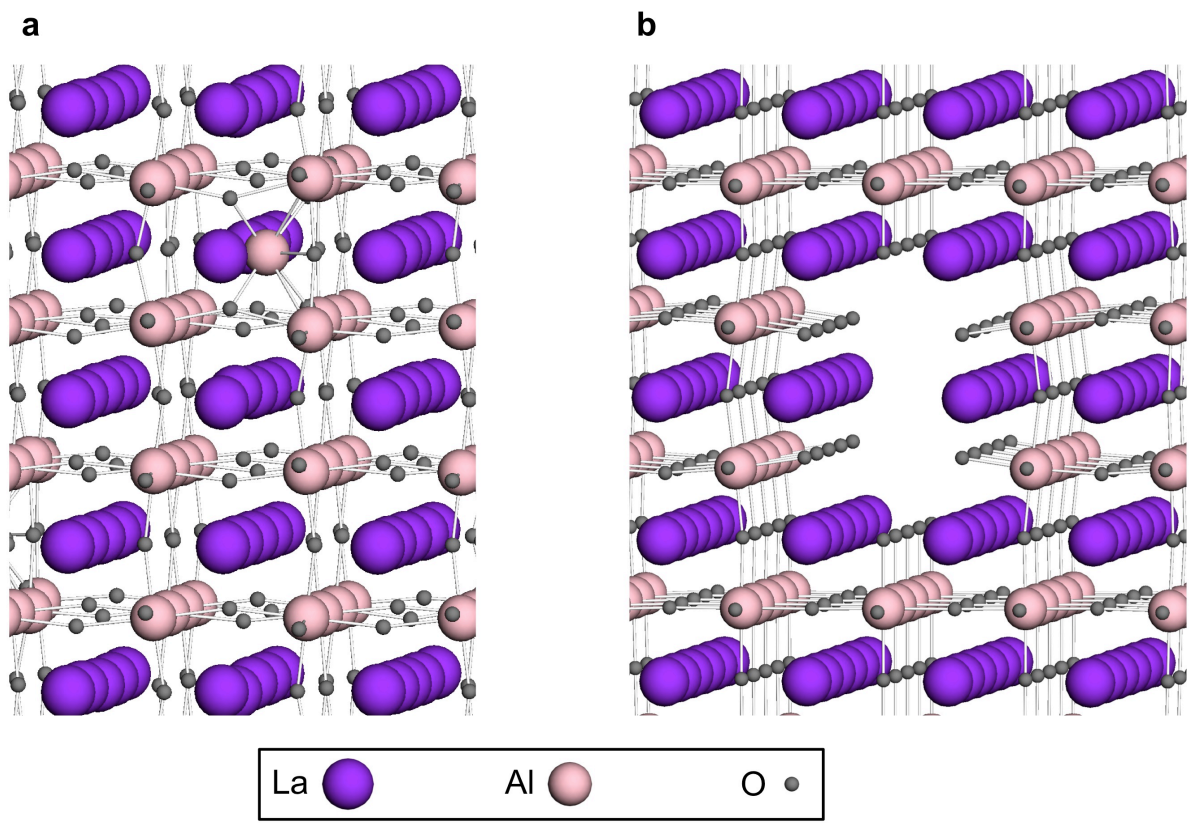

**Figure S13. Different views of lowest energy structures determined with density functional theory. a,** In Al-rich films, Al substitutes for La and shifts off center. The lowest-energy structure is shown as viewed slightly tilted from the [100] direction. **b,** In La-rich films, $Al_2O_3$-vacancy-complexes form, which are periodic in the [001] direction. The smallest $Al_2O_3$-vacancy-complex is shown as viewed slightly tilted from the [001] direction.

**Supplementary Discussion 4.**
**Cation composition mapping using STEM-EELS**

STEM-EELS spectroscopic images were acquired on samples 2–4A and 2–4B, representative La-rich and Al-rich films, grown on pieces of SrTiO$_3$ cut from the same initial substrate. The Ti-$L_{2,3}$, La-$M_{4,5}$ and Al-$K$ edges were acquired simultaneously to compute the total concentration of titanium and aluminum on the $B$-site. Due to limitations in the number of channels accessible on the EELS spectrometer, this was immediately followed by an acquisition of the La-$M_{4,5}$, Al-$K$ and Sr-$L_{2,3}$ edges from a neighboring region on the TEM wedge to determine the concentration of lanthanum and strontium on the $A$-site.

The concentration of the cations was determined by performing a power law background subtraction and integration over approximately 10 eV. EELS scattering cross-sections are only accurate to about 30% for comparison between different shells, and thus could not be used to normalize the signal with sufficient precision for this study. The strontium and titanium signals were normalized from signal in the SrTiO$_3$ substrate sufficiently far from the interface. The lanthanum and aluminum signals were normalized from the signal in the LaAlO$_3$ film; all signals were normalized to "1" regardless of the composition of the film. While this normalization cannot be used to determine total concentration, it does preserve the presence or absence of local, relative variations in concentration such as the local dip in $B$-site concentration observed for sample 2–4A. As shown in Fig. S14, if there is no concentration dip at the interface, the normalization cannot introduce one. We also tested the normalized data to the RBS concentrations from the partner set of films, and found the same trends.

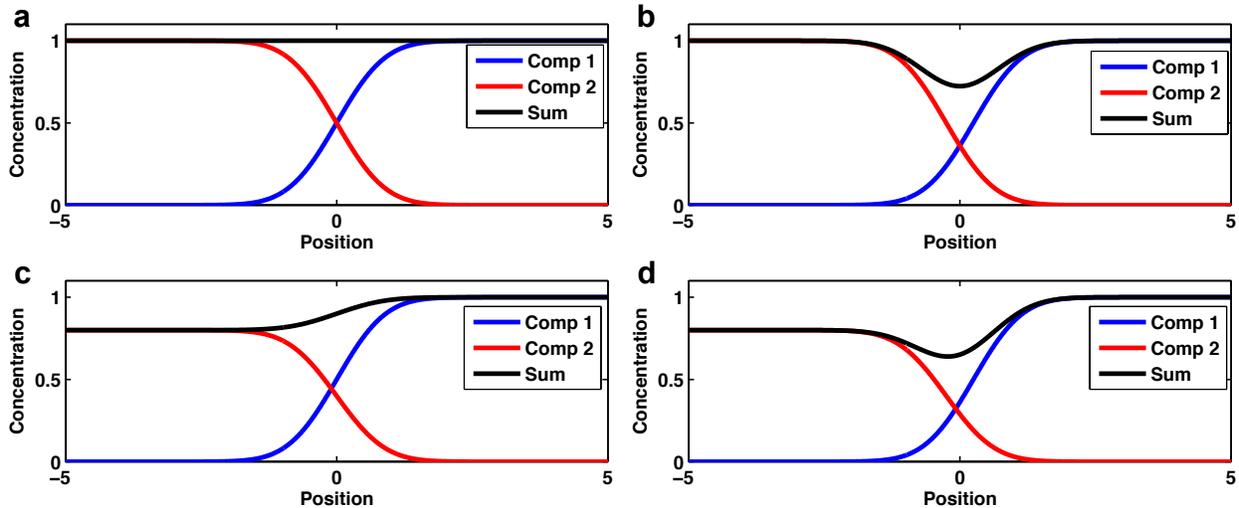

**Figure S14. Simulation to confirm normalization. a,** A simulation of a vacancy free interface with interdiffusion (error function line shape). The total sum of the two components is unity throughout the interface. **b,** Simulation of two normalized components, however with a 25% reduction in the total atomic concentration at the interface. A clear dip in the total concentration is observed, similar to Fig. 4h. **c,** Simulation of two components, one of which has 80% occupancy but free of additional vacancies at the interface. There is no local dip in the total concentration at the interface below the mean values on either side. **d,** Simulation of two components, one of which has 80% occupancy and additional vacancies added at the interface. Once again a local drop in the total atomic profile is seen at the interface.

## Supplementary Discussion 5.
## Oxygen vacancies in SrTiO$_3$ due to growth conditions

The orders of magnitude lower oxygen partial pressures that are generally used in oxide molecular-beam epitaxy in contrast to pulsed-laser deposition is considered to lead to oxygen vacancies, which are well known to dope SrTiO$_3$ substrates with electrons. This overwhelms phenomena occurring at the La$_{(1-\delta)}$Al$_{(1+\delta)}$O$_3$/SrTiO$_3$ interface if the oxygen content of the sample, including the substrate, is not carefully maintained. The samples studied in this work were grown with a distilled ozone source instead of molecular oxygen. The high oxidation power of ozone in contrast to molecular oxygen enables the growth of fully oxygenated samples at lower ozone partial pressures.

To qualitatively determine the oxygen vacancy content in SrTiO$_3$ substrates once exposed to growth conditions, electrical measurements were carried out on a SrTiO$_3$ substrate that was cut into four, 5 mm × 5 mm pieces, labeled *p*, *q*, *r* and *s*. The four SrTiO$_3$ substrate pieces were first etched and annealed via the same steps used to prepare the SrTiO$_3$ substrates for the samples studied in this work. i.e., following the standard procedure[14] to obtain a TiO$_2$ terminated surface. Using a shadow mask with a large via hole of ~ 4 mm × 4 mm, a 100 Å thick chrome layer followed by a 1300 Å thick gold layer were thermally evaporated (at room temperature in a vacuum of ~1×10$^{-5}$ Torr) on the center of the backside of each of these substrate pieces taking care that the chrome-gold layer did not touch the substrate edges. This chrome-gold layer served two purposes: it enabled radiative heating of the substrates in vacuum and later served as a large-area-contact for electrical measurements.

A sample holder with a 4 mm × 4 mm opening was used to mount the substrate pieces such that they could be radiatively heated from the backside. Substrate pieces *p*, *q* and *r* were consecutively loaded into the MBE chamber and exposed to different conditions while the last SrTiO$_3$ piece *s*, served as a control for comparing the electrical data and remained under atmospheric conditions. Substrate piece *p* was heated inside the MBE chamber to 680 °C in a base vacuum of ~ 6×10$^{-9}$ Torr. Substrate piece *q* was heated to the same growth temperature of 680 °C, but in a molecular oxygen background partial pressure of 1×10$^{-6}$ Torr while substrate piece *r* was similarly heated to 680°C in the presence of ozone providing a background pressure of 1×10$^{-6}$ Torr. In all cases the 680 ºC substrate temperature was sustained for 20 minutes (the typical growth time for the 8 unit cell thick La$_{(1-\delta)}$Al$_{(1+\delta)}$O$_3$ on SrTiO$_3$ films studied in this work) before cooling down. The substrate pieces *q* and *r* were cooled to 200 ºC while maintaining the same oxygen or ozone background pressure at which point the molecular oxygen or ozone flow was closed and the substrates were immediately taken out of the MBE chamber and exposed to atmospheric conditions. Note that substrate piece *r* was exposed to the same conditions as the La$_{(1-\delta)}$Al$_{(1+\delta)}$O$_3$/SrTiO$_3$ samples studied in this work with the absence of opening and closing the source shutters to grow a La$_{(1-\delta)}$Al$_{(1+\delta)}$O$_3$ film.

At this point, using a shadow mask with a via of ~ 4 mm × 4 mm, a 100 Å thick chrome layer followed by a 1300 Å thick gold layer was thermally evaporated as before on the center of the top surface of each of these substrate pieces. The top and bottom metal contacts enabled us to make resistance measurements (perpendicular to the surface) of the SrTiO$_3$ substrate pieces as well as capacitance measurements in a parallel-plate capacitor geometry. Resistance and capacitance measurements made on these samples at room temperature (~295 K) and at 77 K are given in Table S1.

**Table S1. Resistance, capacitance and loss tangent measurements on SrTiO$_3$ substrates exposed to different conditions.**

| Substrate piece | Measurements at 295 K | | | Measurements at 77 K | |
|---|---|---|---|---|---|
| | Resistance | Capacitance (pF) | Loss tangent | Capacitance (pF) | Loss tangent |
| $p$ (vacuum) | < 20 KΩ | - | - | - | - |
| $q$ (O$_2$) | > 10 GΩ | 46.1 | 0.0013 | 241.2 | 0.0007 |
| $r$ (O$_3$) | > 10 GΩ | 57.8 | 0.0005 | 334.3 | 0.0002 |
| $s$ (control) | > 10 GΩ | 46.3 | 0.0002 | 268.2 | 0.0003 |

As expected the substrate piece $p$, which was heated in vacuum was found to be quite conductive (resistance less than 20 KΩ at room temperature) due to loss of oxygen. Capacitance measurements on this substrate were therefore, not feasible. The other substrate pieces were all very insulating with resistances greater than 10 GΩ (the maximum resistance that can be measured with our Keithley nanovoltmeter/current source measurement setup). Capacitance and loss tangent measurements were made at 100 Hz and 1 KHz with an Andeen Hagerling (AH2700A) capacitance bridge on these substrate pieces. As a representative data set, measurements made at 1 KHz are shown in table S1 since both frequencies gave similar capacitance and loss tangent values. The difference in capacitance between the different substrate pieces is mainly due to the ± 15% variation in the surface area of the chrome-gold contacts that form the top and bottom electrodes. The large increase in capacitance from room temperature to 77 K (by approximately a factor of 5) is a result of the increase in the dielectric constant of SrTiO$_3$ as the temperature is lowered.

The loss tangent measurements are proportional to the leakage current that could be due to a presumable loss of oxygen in these substrate pieces. It should be noted that both room temperature and 77 K measurements were made with the substrate pieces mounted on a probe. The long and not completely balanced wires between the substrates and the capacitance bridge were found to set a lower bound for accurate loss tangent values to 0.0005. Values below this should be considered to be within the measurement error. Substrate piece $q$, which was heated in a background pressure of 1×10$^{-6}$ Torr of oxygen was found to have a loss tangent above this measurement threshold both at room temperature and at 77 K, suggesting some possible loss of oxygen. It should be noted here that the oxygen injector in the MBE system is directed straight at the substrate and is located a distance of 64 mm from the substrate, which could lead to a much higher oxygen partial pressure at the substrate than the 1×10$^{-6}$ Torr background pressure measured by an ion gauge. This possibly much higher partial pressure of oxygen at the substrate should be taken into account when considering the small increase in the loss tangent observed for substrate piece $q$ compared to that of the control. The substrate piece $r$, however, which was heated in a background pressure of 1×10$^{-6}$ Torr of distilled ozone had loss tangent values comparable to the control and below the measurement threshold. This data clearly demonstrates that the growth conditions used to grow the La$_{(1-\delta)}$Al$_{(1+\delta)}$O$_3$/(001) SrTiO$_3$ films studied in this work does not lead to an extra conduction mechanism related to a loss of oxygen from the bulk of the SrTiO$_3$ substrates.